\pdfoutput=1
\documentclass[]{JHEP3}

\usepackage{graphicx}

\title{Galaxy rotation curves from General Relativity with Renormalization Group corrections}

\author{Davi C. Rodrigues, Patricio S. Letelier 
\\ Departamento de Matem\'atica Aplicada, IMECC, 
Universidade Estadual de Campinas, 13083-859, Campinas, SP, Brazil 
\\ E-mail: \email{davi@ime.unicamp.br, letelier@ime.unicamp.br}}
\author{Ilya L. Shapiro 
\\ Departamento de F\'isica, ICE , Universidade 
Federal de Juiz de Fora, 36036-330, Juiz de Fora, MG, Brazil. 
\\ E-mail: \email{shapiro@fisica.ufjf.br} }

\abstract{ We consider the application of quantum corrections computed using renormalization group arguments in the astrophysical domain and show that, for the most natural interpretation of the renormalization group scale parameter, a gravitational coupling parameter $G$ varying $10^{-7}$ of its value across a galaxy (which is roughly a variation of  $10^{-12}$ per light-year) is sufficient to generate galaxy rotation curves in  agreement with the observations.
The quality of the resulting fit is similar to the
Isothermal profile quality once both the shape of the
rotation curve and the mass-to-light ratios are considered
for evaluation. In order to perform the analysis, we
use recent high quality data from nine regular disk
galaxies. For the sake of comparison,
the same set of data is modeled also for the Modified
Newtonian Dynamics (MOND) and for the recently proposed
Scalar Tensor Vector Gravity (STVG). At face value, the model based on
quantum corrections clearly leads to better fits than
these two alternative theories.}


\begin{document}

\def\<{\left \langle}
\def\>{\right \rangle}
\def\[{\left\lbrack}
\def\]{\right\rbrack}
\def\({\left(}
\def\){\right)}
\newcommand{\be}{\begin{equation}}
\newcommand{\ee}{\end{equation}}
\newcommand{\ea}{\end{eqnarray}}
\newcommand{\ba}{\begin{eqnarray}}
\newcommand{\prt}{{\partial}}
\newcommand{\diag}{\mbox{diag}}
\newcommand{\tr}{\mbox{tr}}
\newcommand{\grad}{\ensuremath{\vec{\nabla}}}
\newcommand{\bs}{\begin{sideways}}
\newcommand{\es}{\end{sideways}}
\newcommand{\chir}{\chi^2_{\mbox{\tiny{red}}}} 
\newcommand{\Newt}{{\mbox{\tiny Newt}}}
\newcommand{\Iso}{{\mbox{\tiny Iso}}}
\newcommand{\mond}{{\mbox{\tiny MOND}}}
\newcommand{\stvg}{{\mbox{\tiny STVG}}}
\newcommand{\IsoInf}{{\mbox{\tiny Iso} \infty}}
\newcommand{\dg}{\dagger} 

\renewcommand{\vec}[1]{{\bf #1}}
\renewcommand{\Re}{\,\mbox{Re}\,}
\renewcommand{\Im}{\,\mbox{Im}\,}
\def\beq{\begin{eqnarray}}
\def\eeq{\end{eqnarray}}
\def\ln{\,\mbox{ln}\,}
\def\Det{\,\mbox{Det}\,}
\def\det{\,\mbox{det}\,}
\def\tr{\,\mbox{tr}\,}
\def\diag{\,\mbox{diag}\,}
\def\Tr{\,\mbox{Tr}\,}
\def\sTr{\,\mbox{sTr}\,}
\def\Res{\,\mbox{Res}\,}
\renewcommand{\Re}{\,\mbox{Re}\,}
\renewcommand{\Im}{\,\mbox{Im}\,}
\def\lap{\Delta}
\def\sla{\!\!\!\slash}
\def\al{\alpha}
\def\bet{\beta}
\def\ch{\chi}
\def\ga{\gamma}
\def\de{\delta}
\def\vp{\varepsilon}
\def\ep{\epsilon}
\def\ze{\zeta}
\def\io{\iota}
\def\ka{\kappa}
\def\la{\lambda}
\def\na{\nabla}
\def\pa{\partial}
\def\ro{\varrho}
\def\rh{\rho}
\def\si{\sigma}
\def\om{\omega}
\def\ph{\varphi}
\def\ta{\tau}
\def\th{\theta}
\def\te{\vartheta}
\def\up{\upsilon}
\def\Ga{\Gamma}
\def\De{\Delta}
\def\La{\Lambda}
\def\Si{\Sigma}
\def\Om{\Omega}
\def\Te{\Theta}
\def\Th{\Theta}
\def\Up{\Upsilon}

\section{Introduction} \label{intro}

The dynamics of many galaxies looks like as if the main part of 
their masses is distributed in a different way than their light 
is, with  heavier densities than those expected from the emitted 
radiation by the stars and gas. Curiously, although galaxies 
are far from being ``hairless", like fundamental particles or 
black holes, these masses discrepancies follow considerably strong 
patterns, including a series of correlations with the luminous 
matter, e.g. \cite{urc, btf}. 

Usually, in order to solve this missing matter problem, a qualitatively new kind of matter, dark matter (DM), is evoked. Among all kinds of dark matter, current cosmological observations (like the large scale structure and the microwave background anisotropies) favor the collisionless and ``cold'' type of dark matter (CDM). However the CDM paradigm faces some difficulties at galactic scales, like the angular momentum issues on galaxy formation, the discrepancies on the number of satellite galaxies,  and the cuspy dark matter density profile (see \cite{primack} for a recent review). The CDM paradigm has not yet yielded a satisfactory dark matter profile for spiral galaxies from the simulations of cosmological evolution. The problem with the cuspy dark matter distribution in galaxies was already pointed in \cite{flores, moore1994}, and in particular both the Navarro-Frenk-White \cite{nfw} and the Moore \cite{moore} dark matter profiles  have a cusp on the dark matter density at the galactic center whose dynamical consequences do not match the observations; the discrepancies with the observations becoming particularly clear for the Low Surface Brightness galaxies. The problematics associated with this cuspy dark matter profile, and possible solutions, were systematically analyzed in many references, including \cite{blokrubin, spano,  Gentile, ThingsRot, ohthings, Donato}, see \cite{corecuspblok} for a recent review.

It is well known that the parameters of the dark matter profiles that best fit the observed rotation curves  (like the Isothermal \cite{BBS} or the Burkert\footnote{Considering the CDM simulations, the Burkert profile has merits over the Isothermal one, in particular for its large $R$ behavior and its finite total mass; see also \cite{burkertCDM}.  } \cite{burkert} ones, which have a core  and were proposed on phenomenological grounds) display many correlations with the baryonic matter behavior, e.g. \cite{mcgaughdmlsb, corecor, Donato, BBS, burkert}, see also \cite{Cthesis}. These correlations between a galaxy rotation curve and its baryonic matter suggest another interpretation to the missing mass problem in galaxies, namely  a change of the gravitational law itself instead of introducing DM.  This has lead to diverse proposals, including modifications on the law of inertia  \cite{mond} and its extensions to General Relativity (e.g., \cite{revteves}), proposals with extra-dimensions \cite{extrad}, actions of General Relativity with an extra affine connection or a new graviton \cite{ebi}, conformal gravity \cite{Mannheim}, non-symmetric metrics \cite{BrownMoffat}, among others possibilities. MOND \cite{mond, sandersmcgaugh} is by far the most studied example of the last approach. Although it seems to be consistent with  the general features of the galactic dynamics, its concordance with observations (if the original recipe is applied) started to be questioned in the light of 
the higher precision recent observations and the improvements on the  stellar mass-to-light ratios constraints, e.g. \cite{Gentile, m33m31, tevesgalaxy, mondmilky, mondearlydisk}.

One can suppose that the deviation from the gravitational 
(either Einstein or Newton) law is due to the quantum effects,
such as semiclassical corrections, effects of quantum gravity, 
consequences of string theory physics, extra dimensions, branes
or some other (maybe yet unknown) model of ``quantum gravity''.  

\vskip 1mm

The corrections to the Newton law is a relatively common 
feature of the different models of ``quantum gravity'', 
including higher derivative quantum gravity \cite{frts82}, 
different versions of the effective low-energy quantum theory 
of gravitational field \cite{AntMot,don,Wood}, so-called 
non-perturbative quantum gravity based on the hypothesis of 
the existence of the non-Gaussian UV fixed point \cite{Reuter}
and, finally, in the semiclassical approach to quantum gravity
\cite{book} (see also references therein). The application 
of these corrections has been elaborated recently in the 
cosmological \cite{nova,CCfit,Gruni} and astrophysical 
\cite{Gruni} areas. One can see the recent paper \cite{DCCR}
for the detailed discussion of the quantum field theory 
backgrounds of these quantum effects.  
Let us emphasize, from the very beginning, that in the most 
cases the present day state of art in all mentioned 
approaches to quantum gravity does not enable one to really 
calculate the relevant quantum contributions to the Newton law 
in a unique and consistent way. At the same time, we have
many reasons to believe that these quantum corrections 
can be non-zero and, in principle, may have a measurable 
effect. Typically, the theoretical estimate for the 
quantum contribution to the gravitational law involves 
more or less strong arbitrariness, related to the dependence 
of the quantum corrections on some dimensional parameter
(scale parameter in the renormalization group-based 
approaches, for instance) from one side, and to the physical
interpretation of this parameter from the other side.  

In the present paper we will consider a relatively general 
model of quantum corrections which is mainly controlled by 
covariance. Furthermore we will use a new identification of 
the scale parameter $\mu$ in the astrophysical setting
(rotation curves problem), which we believe to be the more 
natural compared to the one \ ($\mu \sim 1/r$) \
considered before in \cite{Bertol,Gruni,Reuter} (see also 
\cite{Mazzitelli-94}). It turns 
out that this new identification of scale leads to the  
surprisingly good result for the rotation curves of the 
galaxies, which is very close in quality to the output of the 
mainstream approach based on the DM paradigm. 

The previous papers on the subject were based on the 
renormalization group equations coming from the 
higher-derivative quantum gravity \cite{Bertol} (unfortunately, 
the proper renormalization group equations \cite{frts82}
are ambiguous in this case, making their application to 
astrophysics doubtful \cite{nova}), on the assumption of 
the Appelquist and Carazzone decoupling of the quantum 
corrections in the low-energy domain \cite{babic,nova,CCfit} 
and on the assumption of the asymptotic safety and existence 
of the non-Gaussian fixed point in the four-dimensional quantum 
gravity \cite{Reuter}. It is remarkable that all those, in fact 
rather different approaches, converge in predicting qualitatively 
similar logarithmic running of the effective Newton constant with 
the scale $\mu$. It was shown in \cite{Gruni} (see also previous 
qualitatively similar consideration in \cite{Bertol}) that this 
logarithmic behaviour, together with the mentioned above 
identification of $\mu^{-1}$ to the distance from the 
center of the galaxy, $r$, can partially explain the main 
features of the rotation curves without invoking the DM. 
The astrophysical applications of \cite{Bertol}, \cite{Gruni} 
and \cite{Reuter} were performed for the point-like model of 
the galaxy, which is not supposed to be realistic. It was 
pointed out in \cite{Gruni} that it would be very interesting 
to make a more detailed investigation, taking the case of a 
plane or thin disk distribution of mass in the galaxy. We 
present the corresponding analysis and show
that it meets certain obstacles at both theoretical and
phenomenological levels. 

In order to make a conclusive consideration of the 
astrophysical application of the renormalization group 
method, one has to ask whether the $\mu \sim 1/r$
identification is the unique possible one, or there are 
some alternative options. One can remember that the 
identification of scale in the cosmological setting is 
a nontrivial issue, which was treated differently by 
different authors in different papers 
\cite{CCcosm,nova,babic,CCfit,Reuter-t,Reuter}. Furthermore,
was an interesting attempt to construct the regular 
scale-fitting procedure \cite{Guberina-scale}. The output 
of this procedure was the most natural identification of 
$\mu$ with the Hubble parameter $H$, which is the energy
characteristic of the external metric leg of the Feynman 
diagram, corresponding to the quantum correction to the 
expansion of the universe. Indeed, 
similar energy characteristic can be constructed in the 
case of the galaxies rotation curves, and it is indeed 
different from $1/r$. Moreover, this 
``right'' setting of the energy scale leads to the 
immediate and dramatic improvement of the phenomenological 
analysis results. After all, we gain a chance to explain 
the rotation curves on the basis of quantum corrections 
to the classical action of gravity, without using DM. 

The paper is organized as follows. In section 2 we discuss very general features of quantum contributions and establish their likely form, which is based on the general covariance and on the proper existence of these contributions, following \cite{Gruni}. As we have already mentioned above, this existence can not be either proved or disproved at the present-day state of art in this area \cite{DCCR} and it is legitimate to apply a phenomenological approach. Let us note that the covariance enables one to put serious restrictions on the possible forms of quantum corrections in the cosmological setting, the astrophysical ones can be deduced starting from the cosmological setting result and the assumption of a unique effective action for gravity.  The main new element in section 2 is the new definition of the energy scale, which is appropriate for the application to the rotation curves problem. In section 3 we present the numerical results of our model applied to the data of nine regular galaxies divided into two samples of data. For comparison purposes, our results are shown together with the numerical results of three other proposals, obtained from exactly the same data and procedures. These three proposals are: dark matter (modeled with the Isothermal profile \cite{BBS}), MOND \cite{mond} and STVG \cite{stvg, BrownMoffat}. The first is the leading candidate for the missing matter problem in galaxies and all the universe altogether, MOND is the paradigmatic model for galaxy rotation curves without dark matter and STVG is a recently proposed model for astrophysics and cosmology without dark matter that also uses the variation of the gravitational coupling parameter $G$ to fit galaxy rotation curves. Finally, in section 4 we draw our Conclusions and discuss the perspectives of the theory of quantum corrections (especially the ones based on the renormalization group) as a potential competitor of the standard $\Lambda$CDM model, at both astrophysical and cosmological scales. 

	
\section{Effective action of gravity and its low-energy behaviour}
\label{rg}

In this section we will discuss the possible form 
of Effective Action of gravity (EA), including the 
contributions coming from quantum matter or from the 
quantization of the proper gravitational field.
The introduction to the gravitational theory at 
quantum level can be found in \cite{birdav,book} 
(see also recent reviews in \cite{PoImpo} and 
\cite{DCCR}). 

\subsection{Covariance arguments and the scale setting}

The EA is a functional which is the analog of the classical 
gravitational action at quantum level. The EA can be seen 
as a classical action plus quantum contributions. 
The finite part of EA is a non-local functional, which 
usually can not be calculated explicitly. 
Let us use some general features of EA to establish a 
possible form of low-energy quantum effects at the 
cosmological and astrophysical scales 
\cite{nova,CCfit,Gruni,PoImpo}. 
Our assumptions include the covariance of EA, and that 
the low-energy gravity should not have other light 
degrees of freedom except the ones of the metric. 
This means the quintessence models or other entities 
of similar origin are completely out of the consideration 
presented below. 

Under the conditions formulated above, the quantum 
contributions can be taken in form of a power series 
in the derivatives of the metric.  
In the cosmological setting, the second order in 
derivatives term means there are only the 
${\cal O}(H^2)$ - like contributions\footnote{Let us 
note that linear in $H$ noncovariant terms are not 
favored also by the analysis of observational data 
\cite{BPSola}.}. According to 
the consideration presented in \cite{Gruni}, 
this means that the quantum corrected vacuum energy 
density $\rho_\La =  \La/(8\pi G)$ and the Newton 
constant satisfy the equations
\beq
\label{System}
\rho_\La &=& C_0 + \frac{3\nu}{4\pi}\,M_P^2\,\mu^2
\,, \quad
\nu = \frac{\si}{12\,\pi}\,\frac{M^2}{M^2_P}
\\
(\rho+\rho_\La)\,d{G} &+& G\,d{\rho_\La}=0\,,
\nonumber
\\
\rho + \rho_\La &=& \frac{3H^2}{8\,\pi\,G}\,,
\nonumber
\eeq
where $\rho$ is the energy density of matter, 
$\mu=H$, while $\nu$ and $M$ are some undefined 
parameters dependent on the unknown quantum corretions. 

The solution for $G=G(H;\nu)$ can be easily found to be
(see also discussion in \cite{fossil})
\beq
\label{GH}
G(H;\nu)=\frac{G_0}{1+\nu\,\ln\left(H^2/H_0^2\right)}\,,
\eeq
where $G(H_0)=G_0\equiv 1/M_P^2$ is the initial value of $G$.
It is interesting to note that the relation (\ref{GH}) 
implies that the Appelquist - Carazzone decoupling does 
not apply to the Newton constant. The possible reasons for 
that has been explained in \cite{Gruni} and we will not 
repeat them here. 

Let us look for a natural extension of the relation (\ref{GH}).  
The presence of an arbitrary dimensional parameter $\mu$ 
in Eq. (\ref{System}) makes the formula (\ref{GH})
much more general than the purely cosmological setting
\cite{Gruni}. Indeed, both sorts 
of hypothetic quantum contributions come from the Feynman
diagrams which have some number of external legs of a
background metric. The internal lines of these diagrams 
may be of the matter fields in case of semiclassical 
approach or also include quantum gravity propagators 
and vertices in case we deal with some model of quantum 
gravity. 

In all cases the external legs of metric are characterized 
by some dimensional parameter, say the typical energy of 
the gravitational field in a given physical situation. 
This energy should be associated with the dimensional 
parameter $\mu$ which we have introduced before. In the 
cosmological setting, as far as the process of our interest 
is the expansion of the universe, the typical energy of 
the gravitational field can be associated with the Hubble 
parameter, so it is natural to set $\mu=H$ 
\cite{Guberina-scale,Gruni}. However, in another physical 
situation the typical energy parameter $\mu$ may 
have another natural identification. 
From this perspective the relation (\ref{GH}) means we can 
expect similar relation in the general case, where the 
Hubble parameter will be replaced by some, yet unknown,
combination of the metric components and their derivatives. 
In the present work we are interested in the application 
of the general relation 
\beq
\label{Gmu}
G(\mu)=\frac{G_0}{1+\nu\,\ln\left(\mu^2/\mu_0^2\right)}\,,
\eeq
to the rotation curves, and therefore are interested 
to use the phenomenologically best and also natural 
choice for $\mu$ in the corresponding setting. In the 
previous papers on the subject 
\cite{Bertol,Bertolami:1995rt,Gruni,Reuter:2007de} the 
choice of the scale was done according to $\mu \sim 1/r$, 
where $r$ is the radius of rotation for a given star. 
It is obvious that this choice of scale is not really 
natural from the quantum field theory viewpoint, because
the inverse distance can not be seen as a parameter which 
characterizes the energy of the gravitational field quanta 
in the external leg of the corresponding Feynman diagrams. 
The inverse distance is a parameter which does not depend directly 
on the intensity of the gravitational field. 

From the observational viewpoint, the above choice of scale leads to an additive constant contribution to the rotation velocity of galaxies \cite{Bertol,Bertolami:1995rt,Gruni,Reuter:2007de}. It is dubious that an additive constant to the circular velocity, which moreover is the same for every galaxy, can remove or alleviate the need of dark matter on the modeling of galaxy rotation curves (see our forthcoming comments, and \cite{saluccicomment} for a criticism on this approach regarding another model with similar phenomenological consequences). On the other hand, it is remarkable that quantum corrections may lead to a tiny variation of $G$ which, in turn, may contribute  to galaxy rotation velocity in an observable way.

From the consideration presented above it is clear that 
a natural parameter to be associated with $\mu$ should 
be something which is directly related to the intensity 
of the gravitational field. Indeed, we can not know 
this intensity beforehand, since it depends on the 
Newton ``constant'', which is supposed to vary. In the 
present paper, in the description of the gravitational 
field at the galaxy scale, we shall choose the Newton 
gravitational potential as the relevant parameter and set
\beq
\frac {\mu}{\mu_0} =  \( \frac{\Phi_{\mbox{\tiny Newt}}}{\Phi_0} \)^\alpha,
\label{muset}
\eeq
where $\Phi_0$ and $\alpha$ are in this paper presented as phenomenological parameters and $\Phi_\Newt$ is the Newtonian potential computed 
with the boundary condition of it being zero at infinity. Due to the logarithm in the $G(\mu)$ expression, the physical consequences of the above setting are essentially insensitive to the value of $\Phi_0$, as long as its value is not {\it several} orders of magnitude different from a typical value of $\Phi_\Newt$ in the system under consideration, which we will always assume. The dimensionless parameter $\alpha$ has an important role, it must be correlated to the mass of the system under consideration. From the rotation velocity formula for this model, to be seen in eq. (\ref{vsvg}), minimum consistence requirements show that $\alpha$ must go to zero when the mass of the system under consideration goes to zero. Considering the application to galaxy rotation curves, the growth of $\alpha$ with the mass is also expected. The ratio $V_\Newt^2 / \Phi_\Newt$ is independent of the total (baryonic) mass of the system, while the (baryonic) Tully-Fisher law \cite{tf, btf} states that the observed rotation velocity at the outer parts of  disk galaxies  (i.e., their non-Newtonian part) increases with the galactic mass. From  eq. (\ref{vsvg}), the latter statements imply that  $\alpha$ must increase with the galactic mass.  We hope that future developments of the theory here presented will unveil the structure of $\alpha$, or justify it as a reasonable effective parameter.

Contrary to Newtonian gravity, the value of the Newtonian 
potential at a given point does play a significant role 
in this approach. This sounds odd from the perspective of 
usual Newtonian gravity, but this is not so 
from the General Relativity viewpoint, since the latter has no 
free zero point of energy. In particular, the Schwarzschild 
solution is not invariant under a constant shift of the 
potential. The last argument shows that in making the 
choice (\ref{muset}) we are implicitly using relativistic 
arguments. In what follows the reader will see that the 
relativistic effects really play an essential role in 
our model.

Solar system and laboratory experiments on gravitation have not yet reported any deviations from (classical) General Relativity. The above setting is also consistent with these  observations at smaller scales if $\alpha$ decreases sufficiently from its value for the Galaxy to its corresponding value for the Solar system\footnote{Since the baryonic mass in our Galaxy is about 10 orders of magnitude greater than the Solar system mass, it is reasonable to assume that for the aforementioned experiments one would find $\mu/\mu_0 \approx 1$, and hence $G(\mu) \approx G_0$.}. This argument only shows the existence of a way for finding compatibility with the known tests of General Relativity, but the final answer on 
the form of quantum correction can be achieved only through the direct calculation of these corrections from one side and the detailed investigation of their phenomenological 
consequences on another one.


\subsection{The dynamics for a slowly varying gravitational coupling parameter}

Let us consider the dynamics of particles on a theory 
with weak variation of the gravitational coupling 
parameter $G$. We will keep in mind the relation (\ref{Gmu}) 
and pay special attention to the scale identification 
(\ref{muset}), but until some point will write formulas
in a most general form, for $G$ close to $G_0$  and weakly dependent on the 
space coordinates $x^i$, namely $G \prt G/ \prt x^i \approx G_0 \prt G/ \prt x^i$. We call it a slowly varying gravitational coupling parameter\footnote{From the $G(\mu)$ expression, one can check that it is indeed a slowly varying function in the mathematical sense, but the main reason for the use of this name comes from the  approximations $G(\mu)$ is expected to satisfy. }.

Our starting point will be the Einstein-Hilbert action
\footnote{We use $ ( - + + + )$ as the space-time signature. 
For clarity, considering the application to rotation curves, 
the velocity of light $c$ will always be explicit.}  
\beq
S = c^4\,\int d^4 x \sqrt{-g}\;
\frac{R}{16\pi\,G} \,,
\label{a1}
\eeq 
where $g$ is the determinant of the metric, $R$ is the Ricci 
scalar, and $G$ is the variable gravitational coupling.  
Since the cosmological constant $\Lambda$ does not play 
an essential role in the rotation curve analysis, we 
simply set it to zero. 

In what follows we will need a nonrelativistic approximation 
to the Einstein equations and, mainly, to the equation of
motion for the test particle in a theory with variable $G$. 
First of all, let us restrict the consideration by a weakly 
varying $G$, such that 
\beq
\label{varG}
G = G_0 + \de G = G_0(1+\ka)
\,,\qquad 
\left|\ka \right| \ll 1\,.
\eeq
In the last equation $G_0$ is the nonperturbed 
constant value of the Newton gravitational coupling. In 
particular, we will see in the next sections that the 
appropriate value of the parameter $\nu$ in Eq.~(\ref{Gmu}) 
is about $10^{-7}$. This is the value which provides a 
good fit of the rotation curves without the addition 
of dark matter and, definitely, it is very small. 
As a consequence we can always keep only the first order 
in the expansion into series in $\nu$. The same will be 
done with the general ratio $\,\ka={\de G}/{G_0}$.

In order to obtain the expression for the 
corrected gravitational potential in the nonrelativistic 
approximation, we can do a straightforward variation of the action in respect to the metric, or simply follow the method of 
\cite{Gruni}. In order to link the metric in the 
variable $G$ case with the standard one, we perform  
a conformal transformation in the action (\ref{a1}), 
according to 
\beq
{\bar g}_{\mu \nu} = \frac{G_0}{G} ~ g_{\mu \nu}
= (1-\ka)g_{\mu \nu}\,. 
\label{conf}
\eeq
It is easy to see that, up to the higher orders
in $\ka$, the metric ${\bar g}_{\mu \nu}$ satisfies 
the usual Einstein equations with constant $G_0$. 
The nonrelativistic limit of the two metrics 
${\bar g}_{\mu \nu}$ and $g_{\mu \nu}$ is, therefore, 
\beq
g_{00}=-1-  \frac {2 \Phi}{c^2} \quad \mbox{and} \quad 
{\bar g}_{00}=-1- \frac{2 \Phi_\Newt}{c^2} \quad\,,
\label{00}
\eeq
where $\Phi_\Newt$ is the usual Newton potential and 
$\Phi$ is yet unknown potential corresponding to the 
solution of the modifies gravitational theory (\ref{a1}). 
It is easy to see that, in view of (\ref{varG}), 
we have 
\beq
g_{00}=-1-\frac {2 \Phi}{c^2} = (1+\ka){\bar g}_{00}
=  (1+\ka)(-1-\frac {2 \Phi_\Newt}{c^2}) \approx 
 -1-\frac {2 \Phi_\Newt}{c^2} - \ka 
\label{00}
\eeq
and, hence, 
\beq
\Phi = \Phi_\Newt + \frac{c^2} 2\,\ka 
= \Phi_\Newt + \frac{c^2 \, \de G}{2\,G_0}\,.
\label{Phi}
\eeq
For the nonrelativistic limit of the modified gravitational 
force we obtain, therefore, 
\beq
-\Phi^{,i} \,=\, - \Phi^{,i}_\Newt \,-\, \frac{c^2 \, G^{,i}}{2\,G_0}\,,
\label{Phi-deri}
\eeq
where we used the relation $\,G^{,i}=(\de G)^{,i}$. In the last formulas and in what follows, all indices after 
a comma denote partial derivatives, while semicolons denote covariant  derivatives.

Another way to arrive at the same formulas is as follows. 
The affine connections which correspond to the two metrics 
are related as
\beq
\label{newgamma}
\bar {\Gamma}^\mu_{ \nu \lambda} 
=  \Gamma^\mu_{\nu \lambda}  
- \frac 1{2 G_{0}} \(  \delta^{\mu}_{\nu} G_{,\lambda} 
+ \delta^{\mu}_{\lambda} G_{,\nu}  
- g_{\nu \lambda}G^{, \mu} \) + O(\ka^{2})\,,
\eeq   
where 
$$
\bar \Gamma^\mu_{\nu \lambda} 
= \frac12  \bar g^{\mu \alpha} \( \bar g_{\nu  \alpha, \lambda} 
+ \bar g_{\lambda  \alpha, \nu} - \bar g_{\nu  \lambda,  \alpha} \)
\quad \mbox{and} \quad 
\Gamma^\mu_{\nu \lambda} 
=  \frac12  g^{\mu \alpha}  \( g_{\nu  \alpha, \lambda} 
+  g_{\lambda  \alpha, \nu} -  g_{\nu  \lambda,  \alpha}  \)\,.
$$

Since a particle of mass $m$ couples to gravity through the 
action
\beq
- m c \int \sqrt{- g_{\mu \nu} \,dx^\mu \,dx^\nu},  
\label{particleA}
\eeq
the gravitational acceleration felt by a nonrelativistic 
test particle in the static weak field limit is
\beq
\frac{d^2 x^i}{d \tau^2} = - \Gamma^i_{\mu \nu} 
\frac{ dx^\mu}{d \tau} \frac {dx^\nu}{d \tau} \approx 
- \Gamma^i_{00} = - \bar \Gamma^i_{00} - 
\frac 1{2 G_{0}} \( - g_{00} ~ G^{; i} \),
\label{tau}
\eeq
where $\tau$ is the proper time, $d \tau^2 
= - g_{\mu \nu} dx^\mu dx^\nu$, with $x^0 = c t$. 
It is easy to see that the last equation is equivalent 
to (\ref{Phi}). 

Finally, replacing $d\tau^2 = c^2 dt^2$ in (\ref{tau}), we 
arrive at the nonrelativistic equation for the gravitational 
acceleration of a test particle in the form equivalent to
(\ref{Phi-deri}),
\be
\frac{d^2 x^i}{d t^2} \approx - \Phi_\Newt^{,i} 
- \frac {c^2}{2 G_0} G^{,i}\,.
\label{force}
\ee

Let us note that the Newtonian potential in the equation 
above is not exactly the Newtonian potential found from 
$T_{00}$ of the original theory. However, it is easy to 
see that the difference is very small for a very small 
value of $\ka$. 
The relevant energy-momentum tensor is that of a fluid of 
negligible  pressure, namely 	
$$
T_{\mu \nu}^{\mbox{\tiny dust}} = \rho^{\mbox{\tiny dust}}
\; U_\mu U_\nu\,, 
$$ 
where $U^\mu = dx^\mu / d \tau$ is the four-velocity with  
$d \tau^2 = - g_{\mu \nu} dx^\mu dx^\nu$. Furthermore
we know that $\rho^{\mbox{\tiny dust}}$ scales with the 
conformal transformation as $ \propto  (G_0/G )^{3/2}$, 
at least up to the ${\cal O}(\ka)$-terms. 
Hence, 
$$ 
8 \pi G ~ T_{\mu \nu}^{\mbox{\tiny dust}} = 8 \pi G_0 
~ \sqrt{\frac {G}{G_0}} \bar T_{\mu \nu}^{\mbox{\tiny dust}} 
\approx  8 \pi G_0 ~ \bar T_{\mu \nu}^{\mbox{\tiny dust}}\,. 
$$
This approximation is a very good one, since the observational 
uncertainties on $\rho^{\mbox{\tiny dust}}$ are much greater 
than $\,(\ka/2)\,\rho^{\mbox{\tiny dust}}$.  In conclusion, 
within a very good approximation, the Newtonian potential that 
appears in eq.(\ref{force}) is the one induced by the energy 
density of the original theory.

The immediate consequence of the acceleration formula 
(\ref{force}) for an axisymmetric system with radial coordinate $R$ is that the 
rotation velocity $V$ is given by
\beq
V^2 \approx V^2_\Newt  + R ~ \frac {c^2}2  ~ \frac {G_{,R}}{G_0}.
\label{ve}
\eeq
The dimensionless ratio $  \frac {R}2  ~ \frac {G_{,R}}{G_0}$ 
gives the fraction of $c^2$ which should be added to the square 
of the Newtonian circular velocity. Hence, even the tiny variations of 
the gravitational coupling $G$ across a galaxy are sufficient 
to generate significative changes in the observed rotation 
curve.

Fixing the variation of $G$ as in (\ref{Gmu}), the  
relation (\ref{ve}) becomes  
\beq
V^2 = V_{\mbox{\tiny Newt}}^2 -  \nu \;  c^2 ~ R \; \frac {\mu_{, R}}{\mu} \,. 
\label{vemu}
\eeq

Before starting the analysis of the rotation curves on the 
basis of eq. (\ref{vemu}), let us make one more observation
concerning the previous attempts on the $\mu$ scale setting. From eq.(\ref{vemu}), the effect of  setting $\mu \propto R^{- \alpha}$, as suggested in \cite{ Gruni, Reuter:2007de,Bertolami:1995rt} (in these references with $\alpha=1$), is to add a constant contribution proportional to $\alpha$ to the circular velocity. Considering that this new effect has a significant contribution to the rotation curves of galaxies at the farthest observable regions from the galaxy center (large $R$), the least one should expect from this constant is that it should vary from galaxy to galaxy  (in order to replace dark matter). Otherwise, there is no hope to explain the diversity of rotation curve velocities at large $R$ (e.g., see the sample of galaxies studied in this paper). In conclusion, following this approach, $\alpha$ must vary from galaxy to galaxy.  The problem with the setting $\mu \propto R^{- \alpha}$, when applied to real galaxies, appears in the rotation curve behavior close to the galactic center. At this region, there are many galaxies that display a rising curve that starts at few km/s. Hence, if a constant velocity is added to help the large $R$ part  of the rotation curve, the same constant will be troublesome to the central part of the same galaxy.

Another approach to the setting $\mu \propto R^{- \alpha}$ comes from a direct integration of the effective potential of a point particle, instead of using eq. (\ref{vemu}).  First of all, this procedure of summing the potentials constitutes an extra assumption. Second, the integrand has a term that, at most, only depends on the mass trough $\alpha$, thus the meaning of this integration should be analyzed with care. If $\alpha$ has the same value for all the points that constitute the galaxy, then the circular velocity will depend on a new parameter, the model cutoff for very large $R$, which we label $R_f$.  In this case, the resulting rotational velocity is essentially the Newtonian one added by a term proportional to $R^2 / R_f^2 $. Considering the innermost region of a galaxy, this $R^2$ behavior is much better than the previous one in which a constant is added to the Newtonian rotation curve. On the other hand, it is hard to match this quickly increasing behavior to the outer parts of many galaxies. At last, if $\alpha$ depends on the local matter density, its mass dependence for each point should be close to a square root to account for the Tully-Fischer law when galaxies are seen far away (similarly to the approach MOND uses), but this non-linear dependence on the mass is not in conformity with the summation of potentials; therefore this approach also fails. In conclusion, $\mu \propto R^{- \alpha}$ is not a satisfactory setting for real galaxies, both from a theoretical perspective and from a phenomenological one.

Finally, we can conclude that eq. (\ref{muset}) represents 
a more natural dependence 
between the gravitational coupling and the gravitational 
potential in the weak field limit that can account for 
the missing matter problem in galaxies.   

Using the setting of $\mu$ as given by eq. (\ref{muset}) and $V_\infty^2 = \nu \, \alpha \, c^2 $, the  expression for the 
circular velocity becomes simply
\beq
\label{vsvg}
V^2 =  V_{\mbox{\tiny Newt}}^2  
\( 1  - \frac{V_\infty^2}{\Phi_{\mbox{\tiny Newt}} }  \),
\eeq
where the relation $V_\Newt^2 = R \, (\Phi_\Newt)_{,R}$ was used.

For a point particle, 
$V_{\mbox{\tiny Newt}}^2 / \Phi_{\mbox{\tiny Newt}} = -1$, 
hence the above velocity profile describes a rotation curve 
around a massive point particle that is Newtonian in its innermost 
region and becomes flat with velocity $V_\infty$ in its outermost 
region. If matter is the source of gravitational force, than 
$V_\infty$ must have a (direct or effective) mass dependence, 
and in particular it must go to zero when the mass goes to zero. 
Such dependence is also expected from the Tully-Fisher law, since 
it  roughly states that the square of the galaxy rotation curve 
velocity scales with the square root of the galaxy luminosity 
(which in turn is proportional to the galaxy stellar mass).

One may use the Tully-Fisher law, or variations (e.g. \cite{btf}),  
to further constrain the model, possibly unveiling  a suitable 
dependence of $V_\infty$ on other galaxy parameters (e.g.,  mass, morphology, age...). However, 
contrary to the original formulation of MOND, we will let 
$V_\infty$ to be a free parameter which may in the future be fixed.

\section{Galaxy rotation curves} \label{curves}

\subsection{General properties}

The stellar component of many disk galaxies can be  fairly described by single thin disk whose matter density decays exponentially with the radius \cite{freeman},
\be
	\sigma_*(R) =  \frac {M_*}{2 \pi ~ R_D^2}~ e^{- \frac R {R_D} },
	\label{thindisk}
\ee
where $M_*$ stands for the total stellar mass and  $R_D$ is the disk scale length. Usually, a significative  improvement can be found by using a thick disk with matter density given by 
\be
	\rho_*(R,z) = \sigma_*(R) ~ \frac{1}{2 z_0} ~ \mbox{Sech}^2 \left(\frac{z}{z_0}\right),
	\label{thickdisk}
\ee
with $z_0 = R_D / 5$ \cite{z0, ThingsRot}. In the absence of a relevant gas contribution, from the above matter distributions one infers the corresponding Newtonian gravitational potential and hence the velocity profile of the model  (\ref{vsvg}).

In this subsection  the non-Newtonian contribution to the square of the Newtonian velocity, as given by eq. (\ref{vsvg}),
\be
	\label{addsvg}
	 V_\infty^2 ~ R ~ \( \ln |\Phi_\Newt| \)_{,R} ~,
\ee
will be compared to the Isothermal profile contribution. The latter  is one of the phenomenological dark matter velocity profiles that better describes the observed rotation curves (e.g. \cite{Gentile, ThingsRot, Cthesis}), it was studied in many works, and it also has a non-null asymptotic velocity for large galactic radius. The latter property, which is not satisfied by all  dark matter profiles in agreement with observations (e.g., \cite{burkert}), makes the Isothermal profile a natural one to compare with the model here proposed.

The velocity rotation curve of a galaxy modeled with the aid of the Isothermal profile is given by $V^2 = V^2_\Newt + V^2_\Iso$, where
\be
	 V^2_\Iso (R)= \frac{V^2_\IsoInf} {R} \[ R - R_c ~ \mbox{arctan} \( \frac{R}{R_c} \) \]
	\label{viso}
\ee
and $V_\IsoInf^2 = 4 ~\pi~ G_0 ~ \rho_0 ~R_c^2$. This profile is commonly parametrized by $R_c$ and $\rho_0$; but, in this section, since the model previously introduced also has a non-null asymptotic velocity $V_\infty$, it is convenient to parametrize it by $R_c$ and $V_\IsoInf$. The Isothermal profile has no direct dependence on the baryonic matter distribution, on the other hand a number of scaling features that depend on the baryonic matter properties appears when this profile is applied to real galaxies (e.g. \cite{urc, corecor, mcgaughdmlsb, Donato}).  

\FIGURE[l]{\includegraphics[width=7cm]{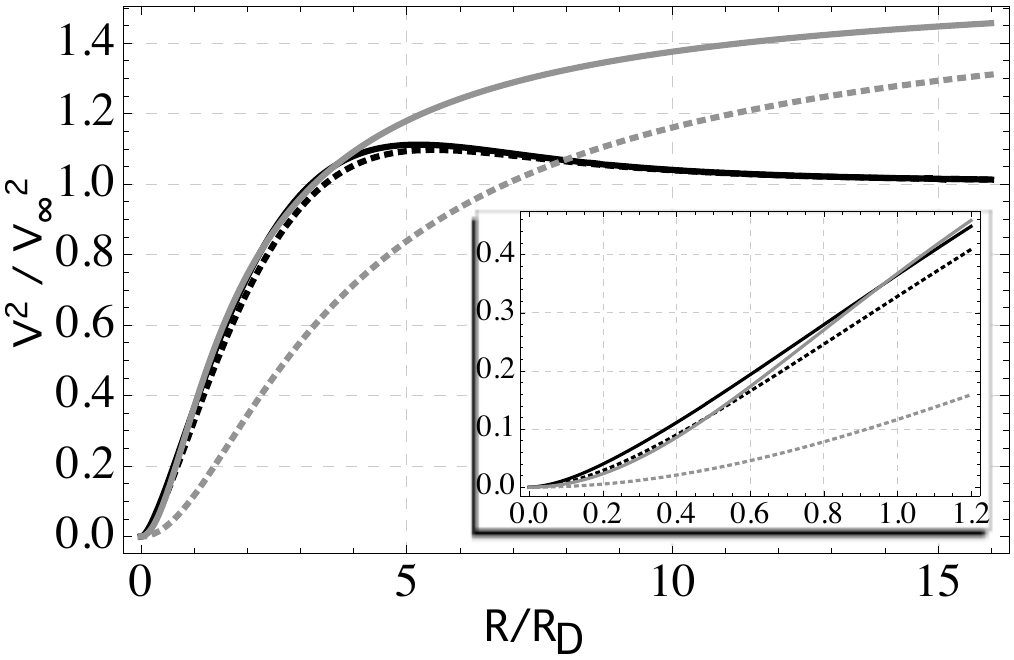} \caption{\label{GenProp1} A comparison between the non-Newtonian contribution induced by our model to the dark matter rotation curve generated by the  Isothermal profile. Black lines: the non-Newtonian square velocity contribution divided by $V_\infty^2$ for thin (solid line) and thick (dotted line) exponential stellar disks. Gray lines: $V^2_\Iso/ V^2_{\infty}$ for $V^2_{\Iso \infty} = 1.6 ~V^2_{\infty}$ with $R_c = 0.95 ~R_D$ (solid line) and $R_c = 2 ~ R_D$ (dotted line). The internal plot is a zoom in the region $R/R_D < 1.2$.}}

In Fig.(\ref{GenProp1}) we plot the dimensionless ratio of the additional contribution (\ref{addsvg}) by $V_\infty^2$, where the Newtonian potential is computed  from both a thin and a thick stellar exponential  disks. In the same figure we  compare the results with the Isothermal profile  contribution, also divided by $V_\infty^2$, for some values of its parameters. Considering a galaxy with negligible gas mass and whose stellar component can be approximated by an exponential disk, Fig.(\ref{GenProp1}) reveals some properties of our model, including:  $i$) the presence of an effective core of radius\footnote{The latter is determined by finding the Isothermal core radius $R_c$ value such that the  $V^2_\Iso$ inflection point happens at the same radius of our model. } $R \approx R_D$;      $ii$) the existence of a rotation curve maximum at $R \approx 5 R_D$, leading to the an approach of the asymptotic velocity from above (contrary to the Isothermal profile); $iii$) the Isothermal profile can be set to be practically identical to the slowly varying $G$ model up to about $5 R_D$; $iv$) the thickness of the stellar disk slightly modifies the stellar rotation curve, which in turn also slightly modifies the non-Newtonian contribution, generating a bigger effective core, and displacing the maximum of the non-Newtonian contribution towards higher radii and lower velocities.

Regarding the item $i$, the correlation between the core radius and the disk scale length is one of the already known correlations between the cored dark matter profiles and the baryonic matter  \cite{corecor}. From the item $ii$ above, a galaxy with negligible gas mass and exponential stellar density profile, whose rotation curve clearly increases for $R \gtrsim 6 R_D$, cannot be satisfactorily described by the the variable gravitational coupling model. To our knowledge, no regular galaxy with these properties have ever been found. Among our sample of galaxies, the galaxy NGC 2841 which can be seen in Fig. (\ref{ngc2841plots}) is the one with the poorest gas content. It does has a relevant bulge component, but its effect is small for large radii. Its rotation curve was measured by more than 10 disk scale lengths, and it clearly displays a decay with the radius in agreement with the above properties. \\[.1in]

Regarding the effects of adding a gas component, Fig. (\ref{GenProp2}) presents an example of the deformation caused by the gas on the non-Newtonian contribution (\ref{addsvg}). Details of the general structure of the gas rotation curve are still under debate (see \cite{tonini} for some possibilities),  but we can fairly illustrate such effect by using a re-escaled version of a ``typical" gas rotation curve, say the one from NGC 3198 \cite{ThingsRot}. To this end it is convenient to work with dimensionless quantities, thus we use the following dimensionless Newtonian potential
$$
	\frac{\Phi_*(R/R_D)}{M_* G_0} +   f ~ 	\frac{\Phi_{\mbox{\tiny gas}}(R/R_D)}{M_{\mbox{\tiny gas}} G_0}, 
$$
where $\Phi_{\mbox{\tiny gas}}$ and  $M_{\mbox{\tiny gas}}$ are deduced from the NGC 3198 data, while $f$ weights the gas influence to the potential. In particular,  $f=1$ represents a galaxy whose stellar mass is equal to its gaseous  mass. Both theoretically and from the numerical result of Fig. (\ref{GenProp2}), one sees that the greater is the gas fraction $f$, the smaller is the non-Newtonian contribution to the inner parts of the galaxy, and the higher is its contribution to the outer parts of the galaxy.  This feature is in agreement with observations, since the gas-rich galaxies are the Low Surface Brightness (LSB) galaxies, which typically display a slowly  increasing rotation curve up to the farthest observed point. 

Figure (\ref{GenProp2}) shows that rotation curves modeled with our model are more susceptible to the gas irregularities than the rotation curves modeled with dark matter (with the Isothermal, Burkert or NFW profiles, for instance). If the gas rotation curve irregularities are too much amplified, this can pose a serious problem to the model, since strong irregularities may be inserted in the resulting rotation curve (see, for example, the ``HI scaling model'' \cite{hoekstra, Gentile} case). Currently it is hard to say if our model amplifies such irregularities either more or less than MOND, since this behavior varies from galaxy to galaxy, depending on a number of factors. The results of our model, to be present in the forthcoming subsection, show that such reflection of the irregularities does not prevent our model to accommodate considerably well the observational rotation curves. It would be interesting to do a tougher investigation on this issue by using a larger sample of galaxies and looking for significative systematic effects.

\FIGURE[l]{\includegraphics[width=7cm]{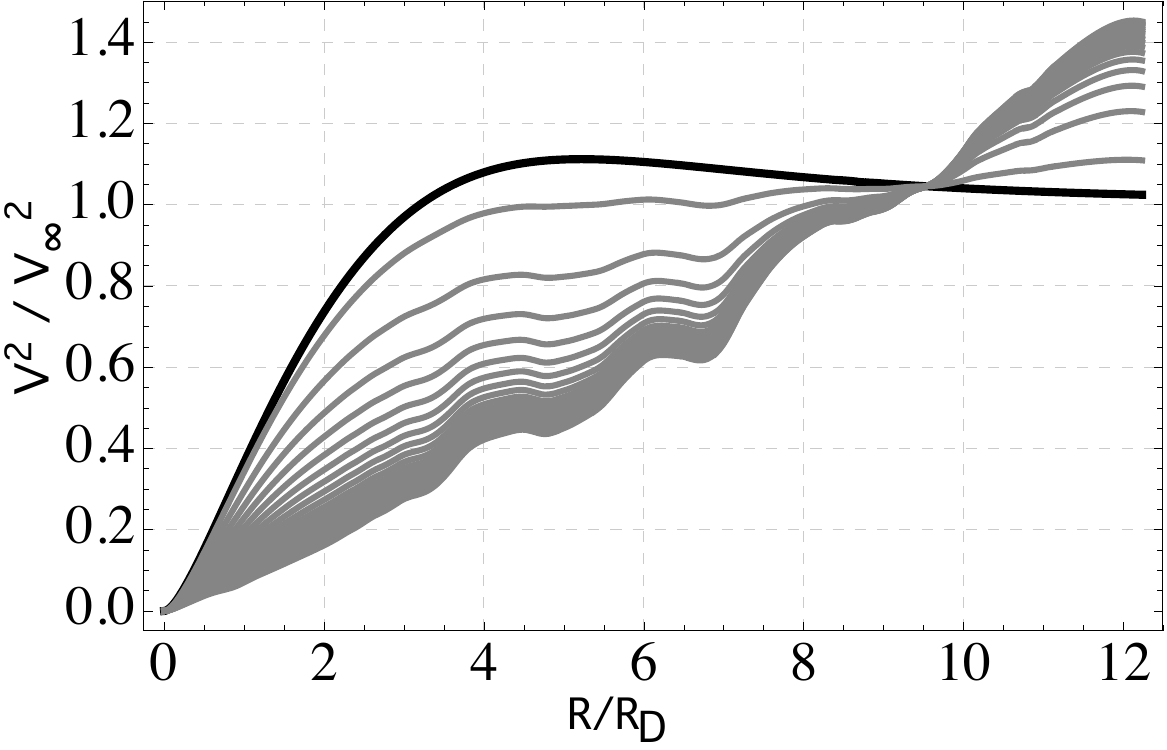} \caption{\label{GenProp2} An illustration of the  deformation on the non-Newtonian contribution induced by the gas presence. Computed using a re-escaled version of the NGC 3198 gas data from THINGS. The solid black line is the squared non-Newtonian velocity contribution divided by $V_\infty^2$,  for a galaxy with thin stellar disk and no gas ($f = 0$). The gray lines are the same as before but with the following values of $f$: 0.2, 0.7, 1.2, 1.7, ..., 9.7.}}


Figure (\ref{GenProp2})  unveils another property of the model. Independently on the value of $f$, all the curves in Fig.(\ref{GenProp2}) pass by a single point. From (\ref{addsvg}), one realizes that the non-Newtonian contribution at a distance  $R_0$, that satisfies $   \( \Phi_* / \Phi_{\mbox{\tiny gas}} \)_{,R} (R_0)= 0$, is not affected by the value of $f$. The existence and position of such point cannot be proved in general due to the diversity of the gas rotation curves. However, in the innermost regions of a galaxy, the ratio  $ \Phi_* / \Phi_{\mbox{\tiny gas}}$ is expected to decrease considerably with the galaxy radius; while, in its outer regions,  the gas contribution typically starts to dominate. The latter leads to an increase in the ratio $ \Phi_* / \Phi_{\mbox{\tiny gas}}$ and thus, under natural assumptions, it implies the existence of the point $R_0$ at the outer parts of many galaxies.

\subsection{Mass decomposition: our results on real galaxies in comparison with other models}

To test the model here presented, which hereafter we label RGGR, we use recent high quality data from nine regular spiral galaxies obtained from two different sources \cite{ThingsRot, Gentile} and compare, for each galaxy, our results with those of three other models, namely: the Isothermal profile, the Modified Newtonian Dynamics (MOND) \cite{mond} and the Scalar-Tensor-Vector Gravity (STVG) \cite{stvg}. The first being a well known and phenomenologically successful profile for dark matter; the second is the most well known, historically important and considerably successful modified Newtonian gravity model; and the third is a recent General Relativity modification that in its weak field limit has similarities with MOND,  while it is also capable of providing good results on lenses and other General Relativity related phenomena (see \cite{revstvg, brownstein} for recent reviews). \\[.1in]

\noindent{\it The models}

{\scriptsize
\TABULAR[c]{l c c}{ 	
	\multicolumn{3}{c}{\bf The free parameters of each model}\\
	\hline \hline \\
	\multicolumn{1}{c}{Model} 	& Model's free 	& Fitting's free \\
	\multicolumn{1}{c}{} 	& parameters 	& parameters \\
	\multicolumn{1}{c}{(1)} 		& (2) 	& (3) 	 \\ 
	\hline \\
	RGGR 					&  $\alpha$ 	&  $\alpha, M_D$ (and $M_B $) 	\\
	Iso			& $\rho_0, R_c$ 	& $\rho_0, R_c, M_D$ (and $M_B $) 		\\
	MOND 				& ------ 	&  $M_D$ (and $M_B $) 		\\
	STVG 				& ------ 	&  $M_D$ (and $M_B $) 		\\ 
	\hline  \label{parameters}}{ \scriptsize (1): ``RGGR'' labels the model we here propose based on Renormalization Group corrections to General Relativity, ``Iso'' labels the dark matter Isothermal  profile, ``MOND'' the Modified Newtonian Dynamics and ``STVG'' the Scalar-Tensor-Vector Gravity model. (2): Lists the model's parameters that are free to vary from galaxy to galaxy. (3): Lists all the parameters whose values are deduced from the fitting procedure. $M_D$ and $M_B$ label the disk and bulge stellar masses respectively. For galaxies with no bulge,  $M_B$ is not a free parameter.} }
	
According to our model, galaxy rotation curves can be explained without the need of dark matter, once renormalization group corrections to General Relativity are taken into consideration. Besides dark matter profiles, a natural model to compare our results with is MOND, since it is currently the most influent and perhaps the most successful approach to explain the missing matter problem in galaxies without dark matter. MOND will be here tested only in its original formulation \cite{mond}, and possible improvements on its results due to small changes in either galaxy inclination or  the distance will not be considered. The details  of such changes to MOND can be found  in \cite{mcgaughmondlsbfit, tevesgalaxy}. To allow for a comparison of model results using the same data and procedures, MOND will be here fitted considering the data from Sample A \cite{ThingsRot} (this is the first time that  MOND is fitted with this data). MONDian fits using the data of Sample B \cite{Gentile} were already done, see \cite{Gentile, tevesgalaxy} for instance. Nevertheless, we will do all the procedures of the fits for MOND to this Sample too, in order to have uniformity on the results and to allow a fair comparison of the results. We remark, in particular, that in the original work on the subject \cite{Gentile} the minimization procedure is different to the one employed here (we use the most usual convention for $\chi^2$, soon to be specified).

The MONDian acceleration, already written in terms of its circular velocity, is given by 
\be
	\frac{V^2_\mond } R ~ \mu \( \frac{V^2_\mond } {R ~ a_0} \) = \frac{V^2_\Newt } R,
\ee
with $\mu(x) = x / \sqrt{ 1 + x^2}$ and $a_0 = 1.2 \times 10^{-8}$ cm/$s^2$ \cite{mcgaughsanders}. Consequently, given the function $V_\Newt(R)$, $V_\mond(R)$ is at once found. 

Likewise the model we present in this paper (RGGR), the STVG model also uses the running of the gravitational coupling motivated by a renormalization group approach to gravity. Its consequences were studied for a large group of galaxies in\footnote{Reference \cite{BrownMoffat} centers its analysis on the Metric-Skew-Tensor-Gravity (MSTG) model, which nevertheless induces the same acceleration law for weak gravitational fields (see also \cite{stvg}). MSTG  also uses the running of $G$ to explain galaxy dynamics. Since here we only analyze galaxy rotation curves, we could label this model to be tested as either STVG or MSTG. --- See however ``Note added''.} \cite{BrownMoffat}. The STVG model is a natural model for us to compare with. To this end, this model will be fitted with the same data and procedures we use to fit our proposed model; in particular we will use the data from Samples A \cite{ThingsRot} and B \cite{Gentile} (this is the first time that this model is fitted with this data).

 The circular velocity induced by  the STVG model is \cite{BrownMoffat}
\be
	V^2_\stvg = V^2_\Newt \[ 1 + \sqrt{ \frac {M_0}{ M_{\mbox{\tiny Total} }}}  \( 1 - e^{-R/r_0} \(1 + \frac R {r_0}\) \) \],
\ee
 where $M_{\mbox{\tiny Total}} $ is the total ``baryonic" mass of the galaxy given by $M_* + M_{\mbox{\tiny gas}}$,    $M_0$ and $r_0$ are related by $M_0 = r_0^2 ~a_0^\stvg / G_0$, with $a_0^\stvg = 6.90 \times 10^{-8}$ cm/$s^2$. For galaxies whose  outermost measured velocity lies at a radius greater than 12 kpc, $r_0$ is set to be $13.92$ kpc; otherwise,  $r_0$ is  $6.96$ kpc \cite{BrownMoffat}. From the theory, $r_0$ is supposed to vary from galaxy to galaxy following the renormalization group flow, but these two values are sufficient for a good agreement  \cite{BrownMoffat}. 
 
 	For the Isothermal profile, we use the velocity given in eq. (\ref{viso}) parametrized by $R_c$ and $\rho_0$, as usual. And, for the RGGR model, we use the velocity given by eq. (\ref{vsvg}), but parametrized by $\alpha$, where $V_\infty =  \alpha ~ c^2 ~10^{-7}$; that is, for convenience, $\nu$ is fixed to be $10^{-7}$. In the end, the parameter that matters  for galaxy rotation curves is the product $\alpha  \nu$.  \\[.1in]
 
	Table \ref{parameters} lists the models and its corresponding free parameters. Besides the parameters that are intrinsic to each model, the disk and bulge stellar masses also appear as free parameters. This is clarified in the next subsection.\\[.1in]

\noindent{\it Stellar masses and the fitting procedure}

To derive the theoretical rotation curve for each galaxy and each model, there is a number of relevant parameters whose values are not known beforehand. For instance, regarding the Isothermal profile, each galaxy can have its own value of core radius $R_c$ and central density $\rho_0$, while for the RGGR model the parameter $\alpha$ is free.  These are parameters that are model specific. Moreover, although the stellar mass distribution of regular disk galaxies is expected to follow approximately an exponential profile \cite{freeman},  the total stellar mass is subject to a considerable uncertainty. The main trend is an increase of the stellar mass with the increase of the absolute luminosity of the galaxy, hence we use the stellar mass-to-light ratio in place of the stellar mass as a  galaxy parameter. One can always convert the latter into stellar mass by using the luminosity values that are provided in Table \ref{initialdata}.

We use the standard procedure for handling the stellar mass-to-light ratio. In a first step,  it is considered to be a free parameter whose value can be found in conjunction with the selected model by a least-square fit, in which the observational velocity and its error bars are considered. Namely, a numerical search for the global minimum of 
\be
	\chi^2 = \sum_{i}^{\mbox{\tiny All observed points}} \( \frac {V^{\mbox{\tiny Obs}}_i - V_{\mbox{\tiny Model}}(R_i) }{\mbox{Error}_i } \)^2
\ee
is done for each  galaxy and each model. In the above, $i$ runs all the observed points of a given galaxy, $V^{\mbox{\tiny Obs}}_i$ and $\mbox{Error}_i$ are respectively the $i$-th observed rotational velocity at the radius $R_i$ and its error. The quality of the rotation curve fit (considering its shape alone) is characterized by $\chi^2_{\mbox{\tiny red}} = \chi^2_{\mbox{\tiny Min}}/(\mbox{\small Number of observed points} - \mbox{\small Number of parameters being fitted})$, where $\chi^2_{\mbox{\tiny Min}}$ is the minimum value of $\chi^2$ found in the previous procedure. This procedure yields a value to the stellar mass (and hence to the stellar mass-to-light ratio) and to additional model parameters if present. The associated 1 $\sigma$ errors to the fitted parameters are deduced by searching for both the maximum and minimum of the same parameter inside the region $\chi^2 \le  \chi^2_{\mbox{\tiny Min}} + 1$, where all other parameters are free.

	The galactic stellar mass is not a true free parameter, since estimates for its value can be derived from stellar population synthesis models, e.g. \cite{belljong, kroupa}. After the fitting procedure, it is analyzed if the resulting mass-to-light ratio is a satisfactory one or not. The expected values are listed in Table \ref{initialdata}.\\[.1in]
	
\noindent{\it About the selected samples}	
	
	In order to analyze  the RGGR model, and to compare it to other proposals, we selected two recent high quality  samples of spiral galaxy data. The data used  can be found in\cite{ThingsRot} (Sample A) and \cite{Gentile} (Sample B). Not all galaxies of these references are here fitted. From the sample of galaxies fitted in \cite{ThingsRot}, we select five whose observed rotation curve are most regular  and  that the stellar component can be very well approximated by one or two thick exponential disks, each disk with density given by eq. (\ref{thickdisk}). Following the conventions of \cite{ThingsRot}, if two exponential disks are used, the disk that dominates the inner part of the galaxy it is called ``bulge". The standard modeling of the bulge with the $R^{1/4}$ profile seems less suitable for this sample of galaxies \cite{ThingsRot}.

	The regularity of the observed rotation curves is an important issue.  Upon avoiding galaxies with either ill understood or just exotic features, emphasis is given to those that the presented  models (built under assumptions of regularity) have a chance of handling and providing meaningful answers.    

	 The same procedures used in  \cite{ThingsRot} for the fitting we also employ here, with the single exception that we use thick exponential disks as approximations for the complete stellar density (instead of only large radii continuations from the photometric data). The exponential disk approximation is a robust one \cite{freeman}. The details on the stellar rotation curve can either improve the concordance with observations (e.g., NGC 2403) or worsen it (e.g., NGC 3198).
	 
	 For the Sample B we use all the conventions used in \cite{Gentile}, apart from their $\chi^2$ minimization. From the five galaxies presented in \cite{Gentile}, only one (ESO 79-G14) is not fitted here since it does not have  an  exponential stellar profile.

Table \ref{initialdata} displays general information on the galaxies that are part either of Sample A or B. The corresponding values of $R_D$ and $R_B$ explicitly cited in \cite{ThingsRot} are not corrected for inclination, and, when cited, refer to the exponential continuation only (except for DDO 154). The values displayed in this table were found from fitting each stellar component rotation curve with  the rotation curve of a thick exponential disk, in which $z_0$ is set to be 1/5 of the corresponding exponential disk scale length. The quoted values of the mass-to-stellar ratio are only used to compare with the values found from the fitting procedure. \\[.1in]
	
	{\scriptsize
\TABULAR[c]{l c c c c c c c}{ 	
	\multicolumn{8}{c}{}\\
	\hline \hline \\
	\multicolumn{1}{c}{Name} 	& $D$ 	& $L_{(D)} $ 	& $L_{(B)}$ 	& $R_D$ 	& $R_{B}$&$ \< Y_{* D}\>$ & $\< Y_{*B}\>$ \\
	\multicolumn{1}{c}{(1)} 		& (2) 	& (3) 		& (4) 		& (5) 	& (6) 	& (7) 		& (8) \\ 
	\hline \\
	\multicolumn{8}{c}{ \it . . . S a m p l e  $ ~  ~$ A . . .}\\[.1in]
	DDO 154 					&  4.3 	&  $0.0082 $ 	& . . . 		& 1.00	& . . .  	& 0.23 		&. . . \\
	NGC 2403 				& 3.2 	&  $1.2$ 		&  0.071		& 1.55 	&  0.410 	& 0.26		& 0.43	\\
	NGC 2841 				& 14.1 	&  $15$ 		&  3.0		& 3.85   	&  0.704 	& 0.53		& 0.60	\\
	NGC 3198 				& 13.8 	&  $3.1$ 		&  . . .		& 3.30 	& . . . 	& 0.57		&. . . 	\\ 
	NGC 3621				& 6.6	& 3.3		& . . .		& 2.09	&. . .		& 0.42		&. . . \\[0.2in]
	\multicolumn{8}{c}{\it . . .  S a m p l e  $ ~  ~$  B . . . }\\[.1in]
	ESO 116-G12 				& 15.3 	& 0.48 		& . . .		& 1.69	& . . . 	& 0.5-1.8		& . . . \\ 
	ESO 287-G13				& 35.6	& 2.3		& . . .		& 3.28	& . . . 	& 0.5-1.8		& . . . \\
	NGC 1090				& 36.4	& 2.5		& . . . 		& 3.41	& . . .	& 0.5-1.8		& . . .\\
	NGC 7339				& 17.8	&0.83		&. . .			& 1.53	&. . . 	& 0.5-1.8		& . . .\\
	\hline  \label{initialdata}}{\scriptsize (2): Distance of the galaxy (Mpc). (3): Total luminosity ($10^{10} L_{\odot}$), the band depends on the sample: 3.6 $\mu$m for Sample $A$ and $I$ band for Sample $B$. (4): ``Bulge" luminosity ($10^{10} L_{\odot}$). (5): Stellar disk scale length (kpc). (6): The ``bulge" scale length (kpc). (7): Expected stellar mass-to-light ratio for the disk. The values quoted for the Sample $A$ are the Kroupa ones. (8): The same as before, but for the `` bulge''.  We stress here that the expected mass-to-light ratios are not used as an input for the fitting procedure, all the fits are done with free stellar masses.} }

\noindent{\it The Newtonian rotation curves}	

	The gas contribution to the Newtonian rotation curve can be evaluated from the Newtonian potential of the gas, which can be straightforwardly  deduced from the gas surface density. (There are different ways to compute the Newtonian rotation curve, but since the RGGR model depends explicitly on the Newtonian potential, this seems to be the most suitable one.) The surface density of the neutral atomic hydrogen (HI) is inferred from the 21 cm radiation of each galaxy (see \cite{walter} for Sample A and \cite{Gentile} for Sample B), and it differs from the gas surface density by a multiplicative constant (below specified). In this work, our starting point for the gas rotation curve is the neutral atomic hydrogen surface density, as provided by the respective collaborations.

	The adopted procedure to find the Newtonian rotation curves goes as follows for Sample A: from the surface density of the neutral hydrogen multiplied by 1.4, to account for the presence of helium and metals, we deduce its Newtonian potential with the boundary condition of it being null at infinity. The latter  is then numerically derived to find the gas rotation curve. For each stellar disk, we use the appropriate disk scale length from Table \ref{initialdata} to find the Newtonian potential of a thick exponential disk, from eq. (\ref{thickdisk}). Except for an overall multiplicative factor for each disk, which is the total stellar mass, the rotation curve is obtained by numerically deriving the potential. 
		
	For Sample B, following the conventions used by the original reference, we use the same procedure as above, except that the surface density of the neutral hydrogen is multiplied by 1.33, to account for the presence of helium, and that the stellar disks are thin; hence the Newtonian potential is found from eq. (\ref{thindisk}).
	
	Some galaxies, close to their center, have a gaseous rotation curve component with a negative contribution to the total circular velocity.  Effectively, this leads to $V^2_{\mbox{\tiny gas}} < 0$ for a certain radius interval, which can describe a physical rotation curve at  that radius if its sum to the other contributions results in a positive total squared circular velocity. In the plots, it is customary to symbolically express  this property in the following way: $ V_{\mbox{\tiny gas}}= \sqrt{V^2_{\mbox{\tiny gas}}}$ if  $V^2_{\mbox{\tiny gas}} >0$, and  $ V_{\mbox{\tiny gas}}= - \sqrt{-V^2_{\mbox{\tiny gas}}}$ otherwise. \\[.1in]

\noindent{\it The results}

The results of the fitting procedure can be found in Tables \ref{resultsA} and \ref{resultsB}, and in the Figs. (\ref{ddo154plots}, \ref{ngc2403plots}, \ref{ngc2841plots}, \ref{ngc3198plots}, \ref{ngc3621plots}, \ref{eso116g12plots}, \ref{eso287g13plots}, \ref{ngc1090plots}, \ref{ngc7339plots}, \ref{ColorVsY_b}).

Comparing the values of $\alpha$ from tables \ref{resultsA} and \ref{resultsB} to the galaxy luminosities in table \ref{initialdata}, one can clearly note a correlation of $\alpha$ with the galaxy luminosity, as already expected from Tully-Fisher arguments. Since the galaxy luminosity correlates with the galactic mass, we see that $\alpha$ increases with the galactic mass, as foretold. Future analysis of this model with a larger sample of galaxies should constrain the dependence of $\alpha$ on the galactic parameters.

Firstly we warn that the absolute values of the $\chi^2_{\mbox{\tiny red}}$ are not of considerable physical significance, since the most relevant information the error bars carry is their relative sizes. There are differences on conventions regarding the error bars values between the Samples A and B \cite{ThingsRot,Gentile}. The relevance of $\chi^2_{\mbox{\tiny red}}$ lies on the comparison of its values between different models for the same galaxy. The RGGR model typically scores almost so well as the Isothermal profile and considerably better than both MOND and STVG. There are however two exceptions, the galaxies NGC 3621 and ESO 116-G12. The RGGR $\chi^2_{\mbox{\tiny red}}$ value is 3 and 4 times greater than the Isothermal ones for the respective galaxies. For NGC 3621, RGGR has a $\chi^2_{\mbox{\tiny red}}$ value as good as STVG, but worse than MOND; while for ESO 116-G12, the RGGR model scores worse than STVG and better than MOND.

The RGGR problem with NGC 3621 lies in the middle radius range, while it can fit well both the rising curve and the irregular shaped final part, reflecting the gas behavior. It is very far from a disastrous result, and, moreover, the result this fit yielded for the stellar mass-to-light ratio is the one closer to the expected Kroupa value. For ESO 116-G12 the problem is not localized, but the final gas wiggle is, for this galaxy, a difficulty.

For the stellar mass-to-light ratios, the model that systematically provides the best values, according to the expected values quoted in Table \ref{initialdata}, is the RGGR model.  

For the Sample A, the RGGR model yielded twice a mass-to-light ratio that is outside the 50$\%$ range  of the expected values. One for DDO 154, a galaxy that no model can yield a good mass-to-light ratio, but that the RGGR model yields the closest one. And another for the bulge of NGC 2841, which is slightly above the allowed range. Probably a very small decrease in the $R_B$ value could solve the previous issue and also improve the rotation curve concordance for small radii.

On comparing our Isothermal results with those of \cite{ThingsRot} care should be taken because we use exponential approximations for the complete stellar components, an approximation that  \cite{ThingsRot} only uses for DDO 154. However, our results are considerably alike. The small structures of the stellar rotation curve can either improve the fit (e.g., NGC 2403), or worsen it (e.g., NGC 3198).

In the Sample B only MOND and STVG found  mass-to-light ratios  outside the expected range, while the Isothermal was three times on the verge of the expected values. The shapes of the curves values found for the Isothermal and MOND models are essentially the same of those found in \cite{Gentile}, but there are systematic differences on the inferred values of the fitted parameters, which are expected since contrary to \cite{Gentile} here we used the standard (or the most common) $\chi^2$ definition for the minimization procedure. 

\vspace{.2in}

The Isothermal profile has parameters strong correlated to be meaningfully fitted in some cases. Two considerably different set of parameters can yield the same $\chi^2_{\mbox{\tiny red}}$ within two precision digits. As an extremum example, we cite an issue we found on NGC 2841. The result that we presented in Table \ref{resultsA} has $\chi^2 = 17.9$, which within three precision digits is not truly the global minima of $\chi^2$, since a completely different  set of parameters resulted into $\chi^2 = 17.7$. The parameters of the latter fit are $Y_{*D} = 1.2 ~ M_\odot / L_\odot$, $Y_{*B} = 1.1 ~ M_\odot / L_\odot$, $R_c = 7.3 $ kpc and $\rho_0 = 25  \times 10^{-3} M_\odot / \mbox{pc}^3$. Note that both stellar mass-to-light ratios have changed from good to  far from the expected values.  The result presented in Table \ref{resultsA} on NGC 2841 was the only one we selected considering both the $\chi^2$ minimization and the likeness of the mass-to-light ratio.

We did some experiences by letting $a_0$ in MOND to be a free parameter. The fittings, considering the shape of the curve, can be significantly enhanced in this picture, but no significant improvement on the mass-to-light ratios was found. For the latter, it is necessary to change the shape of $\mu(x)$, likewise done in \cite{tevesgalaxy}.

{\scriptsize
\TABULAR[c]{l c c c c c  c c }{ 	
	\multicolumn{8}{c}{ \bf Sample A Results }\\
	\hline \hline \\
	\multicolumn{1}{c}{Name}	& Model 	& $\chir$ 	& $Y_{* D}$ 									& $Y_{* B} $ 					& $\alpha$ 			& $R_c$ 				& $\rho_0$ \\
	\multicolumn{1}{c}{(1)} 	& (2) 			& (3) 		& (4) 											& (5) 							& (6) 				& (7) 				& (8) \\ 
	\hline \\
	DDO 154 						&RGGR 			&  0.39 		&2.1 $\pm 0.2^	\dg$						& . . . 							&0.204 $\pm$ 0.006		& . . . 				& . . .\\[.05in]
	DDO 154 						& Iso 			&  0.28 		& ${3.5^{+0.6}_{-0.7}}^\dg$& . . .	 	&. . . 								&$2.7 \pm {0.4}$		& $9 \pm 2$\\[.05in]
	DDO 154 						& MOND 		&  3.6		& ${0.000^{+ 0.005}_{-0.000}}^\dg$& . . .	 							&. . . 				&. . . 				& . . . \\[.05in]
	DDO 154 						& STVG 		&  3.2		& 3.1 $\pm 0.1^\dg$						& . . .	 							&. . . 				&. . . 				& . . . \\[.15in]
	NGC 2403 					& RGGR			& 0.57		& 0.31 $\pm$ 0.01							&  0.40 $\pm 0.03$ 		& 1.66 $\pm$ 0.01  		& . . . 				& . . .\\[.05in]
	NGC 2403						& Iso				& 0.53		&${0.46^{+ 0.06}_{-0.08}}$		&$1.1 \pm  0.1^\dg$		&. . .					&$3.0  \pm 0.4$ 		&$ 40 \pm 10$  \\[.05in]
	NGC 2403						& MOND		&  3.3		&0.96 $\pm 0.01^\dg$ 					& 0.10  $\pm  0.06^\dg$ 	& . . .			& . . .				& . . .  \\[.05in]
	NGC 2403						& STVG		& 3.2			& $1.02 \pm 0.01^\dg$ 					& $0.39 \pm 0.06$  		& . . . 				& . . . 				& . . . \\[.15in]
	NGC 2841						& RGGR			& 0.18		&$0.68 \pm 0.02$							& $1.02 \pm 0.05^\dg$	& $6.7 \pm 0.1$		&. . .					&. . .	\\[.05in]
	NGC 2841						& Iso	*			&0.13		&$0.81 \pm 0.03$							& $0.9 \pm 0.1$				& . . .				& $2.4^{+0.6}_{-0.5}$	& $220 \pm 20$\\[.05in]
	NGC 2841						& MOND		&1.1			&$1.70 \pm 0.02^\dg$					& $0.35 \pm 0.08$			& . . .				&. . .					& . . .\\[.05in]
	NGC 2841						& STVG		&3.6			&$1.11 \pm 0.02^\dg$					& $1.16 \pm 0.06^\dg$	& . . .				&. . .					& . . .\\[.15in]

	NGC 3198						& RGGR			& 0.88		& $0.85 \pm 0.02$							& . . .			& $1.72 \pm 0.03$		&. . . 				&. . . \\[.05in]
	NGC 3198 					&Iso				& 0.46		& $0.93 \pm 0.06^\dg$					& . . .			&. . .					& $4.2^{+0.6}_{-0.3}	$	& $20^{+6}_{-4}$ \\[.05in]
	NGC 3198 					&MOND	 	&5.2			& $0.65 \pm 0.01$							& . . .			& . . .				& . . .				& . . .\\[.05in]
	NGC 3198						&STVG 		&5.8			& $0.777 \pm 0.008$						& . . .			& . . .				& . . .				& . . .\\[.15in]	
	NGC 3621						& RGGR	&1.4		& $0.374 \pm 0.006$		& . . .			& $1.77 \pm 0.02$		&. . .					&. . .\\[.05in]
	NGC 3621						& Iso		&0.47	&$0.60 \pm 0.01$		&. . .				&. . .					& $7.5^{+0.5}_{-0.4}$	&$9^{+0.7}_{-0.8}$\\[.05in]
	NGC 3621						&MOND	&0.52	&$0.510 \pm 0.005$		&. . .				&. . .					&. . .					&. . .\\[.05in]
	NGC 3621						&STVG	&1.4		&$0.554 \pm 0.004$		&. . .				&. . .					&. . .					&. . .\\[.05in]
	\hline  \label{resultsA}}{ \scriptsize (1) Name of the galaxy. (2) Model used to do the mass decomposition. (3) Reduced $\chi^2$ value. (4) Inferred stellar disk mass-to-light ratios in the $3.6 \mu m$ band  ($L_\odot /M_{\odot}$). (5)  The same of the previous item but for the ``bulge" ($L_\odot /M_{\odot}$). (6) Value of the dimensionless RGGR constant  ($V_\infty^2 = \alpha ~ c^2 ~10^{-7}$). (7) The Isothermal profile core radius (kpc). (8) The Isothermal profile density ($10^{-3} M_\odot / \mbox{pc}^3$ ).  A	``$\dagger$" signs a mass-to-light ratio far from its expected value: either $50\%$ greater or smaller than the quoted Kroupa IMF values (considering the fitting errors). (*) There is a significant degeneracy on the parameters of this model for this galaxy that is not reflected in the formal 1 $\sigma$ errors above, see ``{\it The results}''.   }}

{\scriptsize
\TABULAR[c]{l c c c c  c c }{ 	
	\multicolumn{7}{c}{ \bf Sample B Results  }\\
	\hline \hline \\
	\multicolumn{1}{c}{Name}	& Model 	& $\chir$ 	& $Y_{* D}$ 					& $\alpha$ 			& $R_c$ 				& $\rho_0$ \\
	\multicolumn{1}{c}{(1)} 	& (2) 	& (3) 	& (4) 							& (5) 				& (6) 				& (7) \\ 
	\hline \\ 
	ESO 116-G12 			& RGGR 	& 4.2	& $0.49 \pm 0.04$		 			&  $1.00 \pm 0.03$		& . . . 				& . . . \\ [.05in]
	ESO 116-G12 			& Iso 	& 1.1	& $0.5^{+ 0.1}_{-0.2}$	 			&  . . .				& $2.6 \pm 0.6$		& $40^{+ 20}_{-10} $\\ [.05in]
	ESO 116-G12 			& MOND 	& 6.5	& $0.83 \pm 0.03$		  			&  . . .				& . . .				&. . . \\ [.05in]
	ESO 116-G12 			& STVG 	& 1.4	& $0.89 \pm 0.02$		  			&  . . .				& . . .				&. . . \\ [.15in]
	ESO 287-G13			& RGGR	& 1.7	& $1.26 \pm 0.03$		 			& $2.13 \pm 0.05$		& . . . 				&. . .\\[.05in] 
	ESO 287-G13			& Iso		& 1.5	&$0.44 \pm 0.09$		 			& . . .				& $1.6 \pm 0.1$		& $ 210 \pm 30$\\[.05in] 
	ESO 287-G13			& MOND	& 2.2	&$1.56 \pm 0.02$					& . . .				&. . .					& . . . \\[.05in] 
	ESO 287-G13			&STVG	& 11		&$1.44 \pm 0.02$		 			& . . .				&. . .					& . . . \\[.15in] 
	NGC 1090			& RGGR	& 1.3	& $1.31 \pm 0.05$					&$1.88 \pm 0.05$		&. . .					&. . .  \\[.05in]
	NGC 1090			&Iso		& 0.73	&$1.4 \pm 0.1$			 			&. . .					& $4 \pm 1$			& $20^{+20}_{-10}$ \\[.05in]
	NGC 1090			&MOND	&4.3		&$1.19 \pm 0.02$						&. . .					&. . .					&. . .\\[.05in]
	NGC 1090			&STVG	&8.6		&$1.12 \pm 0.02$					&. . .					&. . .					&. . .\\[.15in]
	NGC 7339			& RGGR	&0.96	&$1.34 \pm 0.09$		 			&$1.6 \pm 0.1$			&. . .					&. . .\\[.05in]
	NGC 7339			&Iso		&1.1		&$1.9 \pm 0.1$						&. . .					&$4 \pm 1$			&$40^{+30}_{-10}$\\[.05in]
	NGC 7339			& MOND	& 4.1	&$2.19 \pm 0.03^\dagger$			&. . . 				&. . . 				&. . . \\[.05in]
	NGC 7339			& STVG	& 0.92	&$1.91 \pm 0.02^\dagger $   			&. . . 				&. . . 				&. . . \\	
	\hline  \label{resultsB}}{\scriptsize  (1) Name of the galaxy. (2) Model used to do the mass decomposition. (3) Reduced $\chi^2$ value. (4) Inferred stellar disk mass-to-light ratios in the $3.6 \mu$ band for Sample $A$ and in the $I$ band for Sample $B$ ($L_\odot /M_{\odot}$). (5) Value of the dimensionless RGGR constant, $V_\infty^2 = \alpha ~ c^2 ~10^{-7}$. (6) The Isothermal profile core radius (kpc). (7) The Isothermal profile density ($10^{-3} M_\odot / \mbox{pc}^3$ ). 	``$\dagger$" signs a mass-to-light ratio outside from its expected value (i.e., $Y_*$ outside the range $0.5 - 1.8 ~ L_\odot /M_{\odot}$, considering the 1 $\sigma$ error values). }}

\FIGURE[l]{\includegraphics[width=15cm]{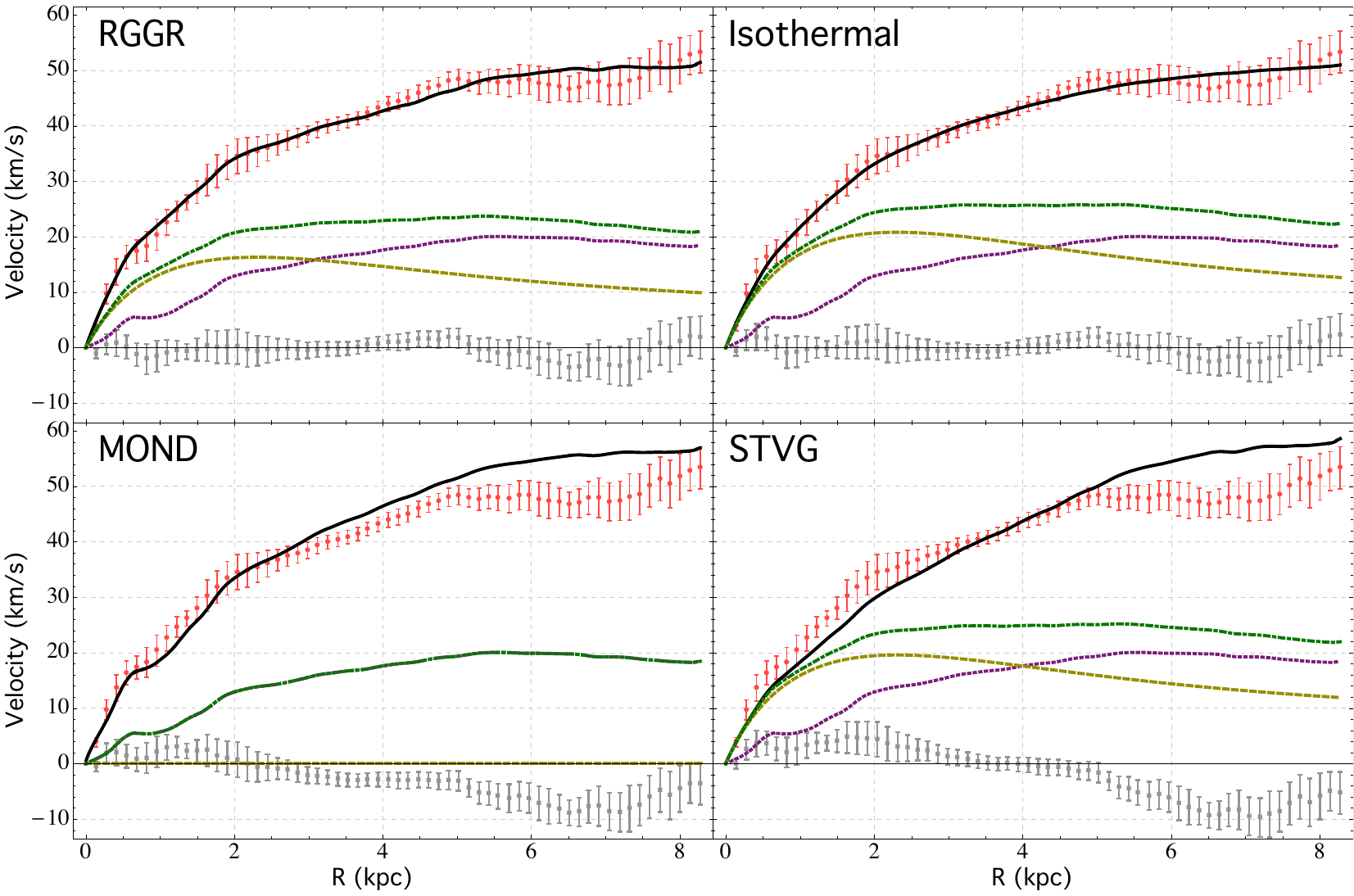} \caption{\label{ddo154plots} DDO 154 rotation curve fits. The red dots and its error bars are the rotation curve observational data, the gray ones close to the abscissa are the residues of the fit. The solid black line for each model is its best fit rotation curve, the dashed yellow curves are the stellar rotation curves from the bulge (if present) and disk components,  the dotted purple curve is the gas rotation curve, and the dot-dashed green curve is the resulting Newtonian, with no dark matter, rotation curve.}}

\FIGURE[l]{\includegraphics[width=15cm]{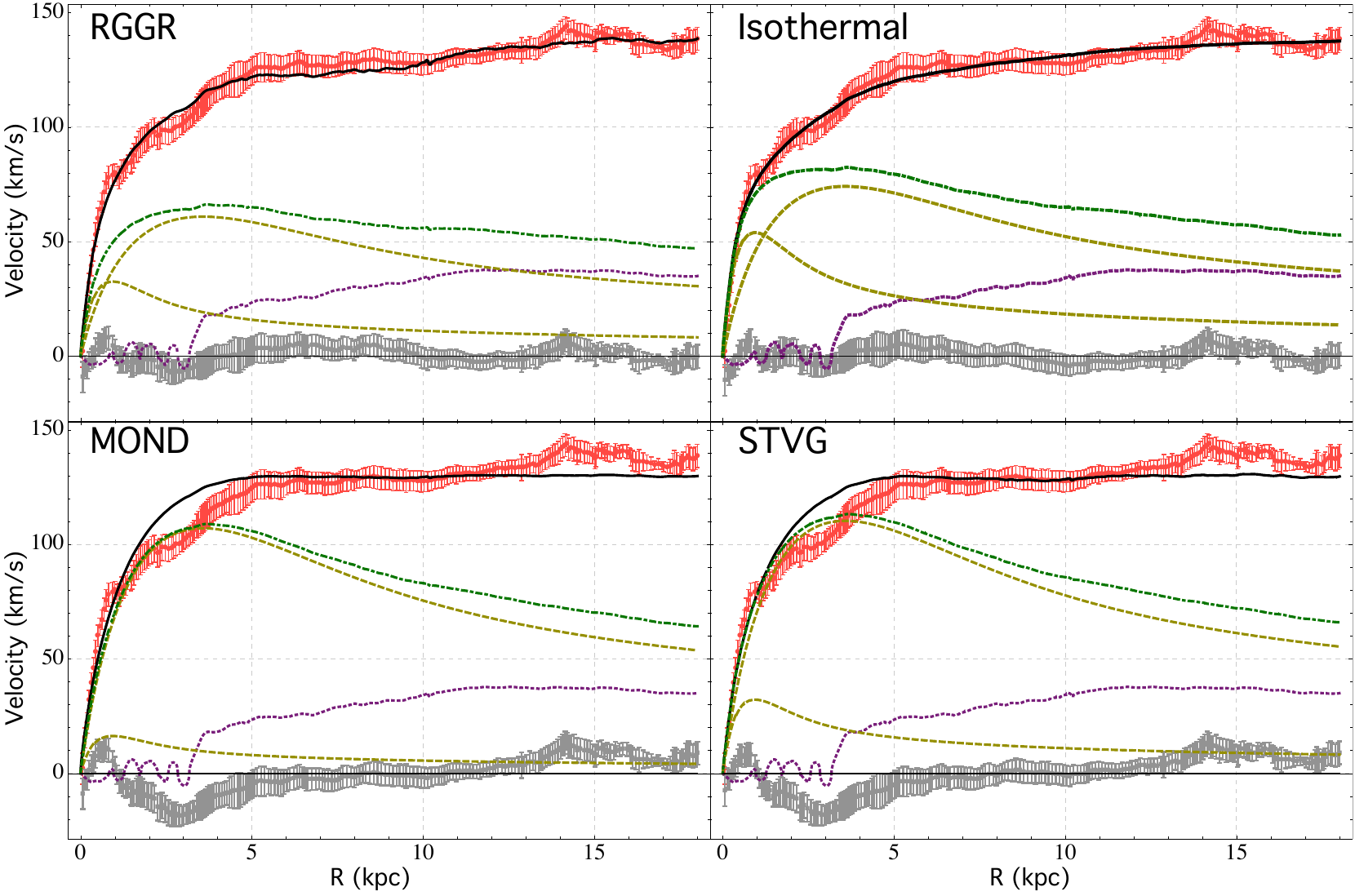} \caption{\label{ngc2403plots} NGC 2403 rotation curve fits. See Fig.(\ref{ddo154plots}) for details.}}

\FIGURE[l]{\includegraphics[width=15cm]{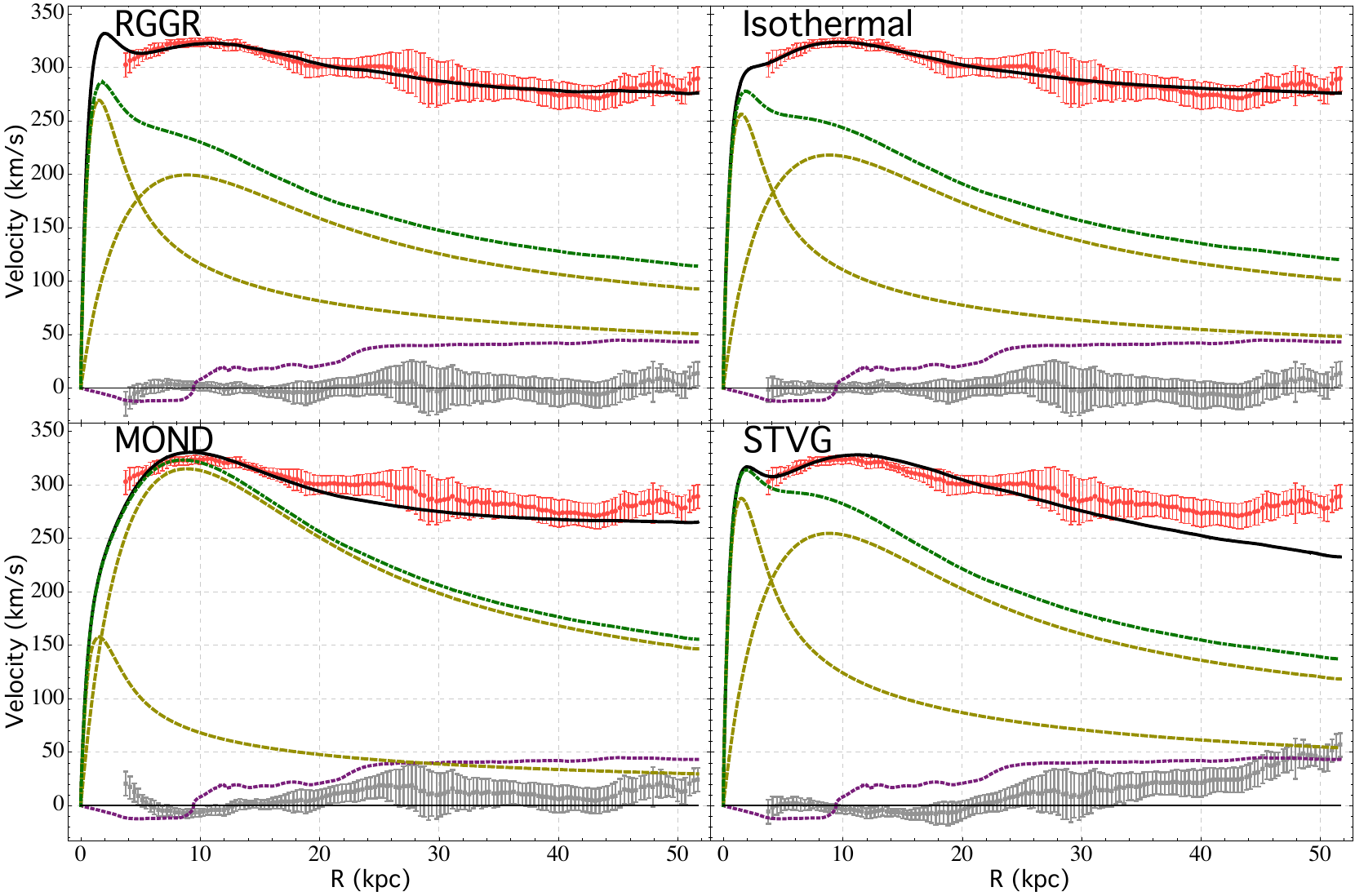} \caption{\label{ngc2841plots} NGC 2841 rotation curve fits. See Fig.(\ref{ddo154plots}) for details.}}

\FIGURE[l]{\includegraphics[width=15cm]{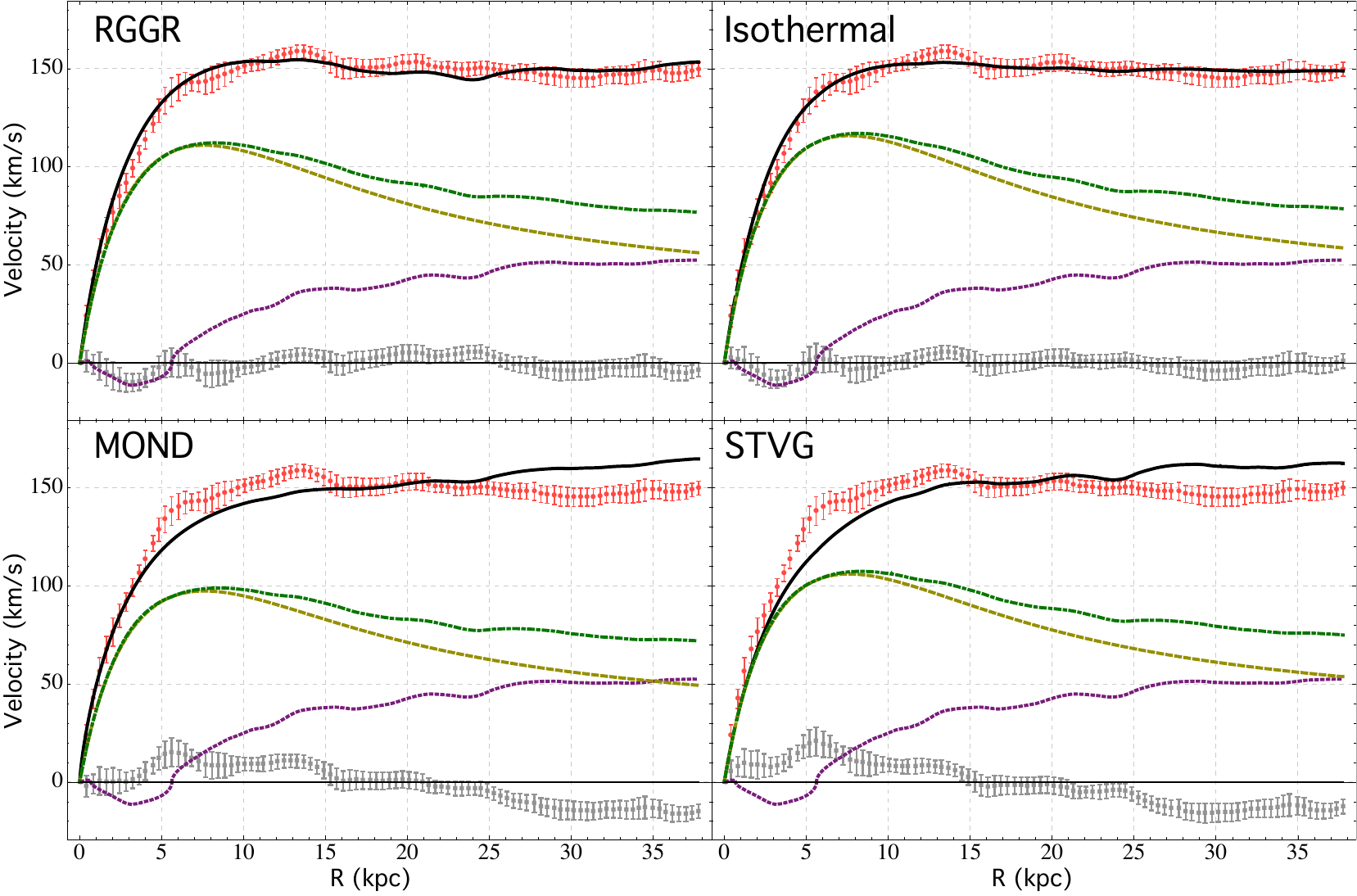} \caption{\label{ngc3198plots} NGC 3198 rotation curve fits. See Fig.(\ref{ddo154plots}) for details.}}

\FIGURE[l]{\includegraphics[width=15cm]{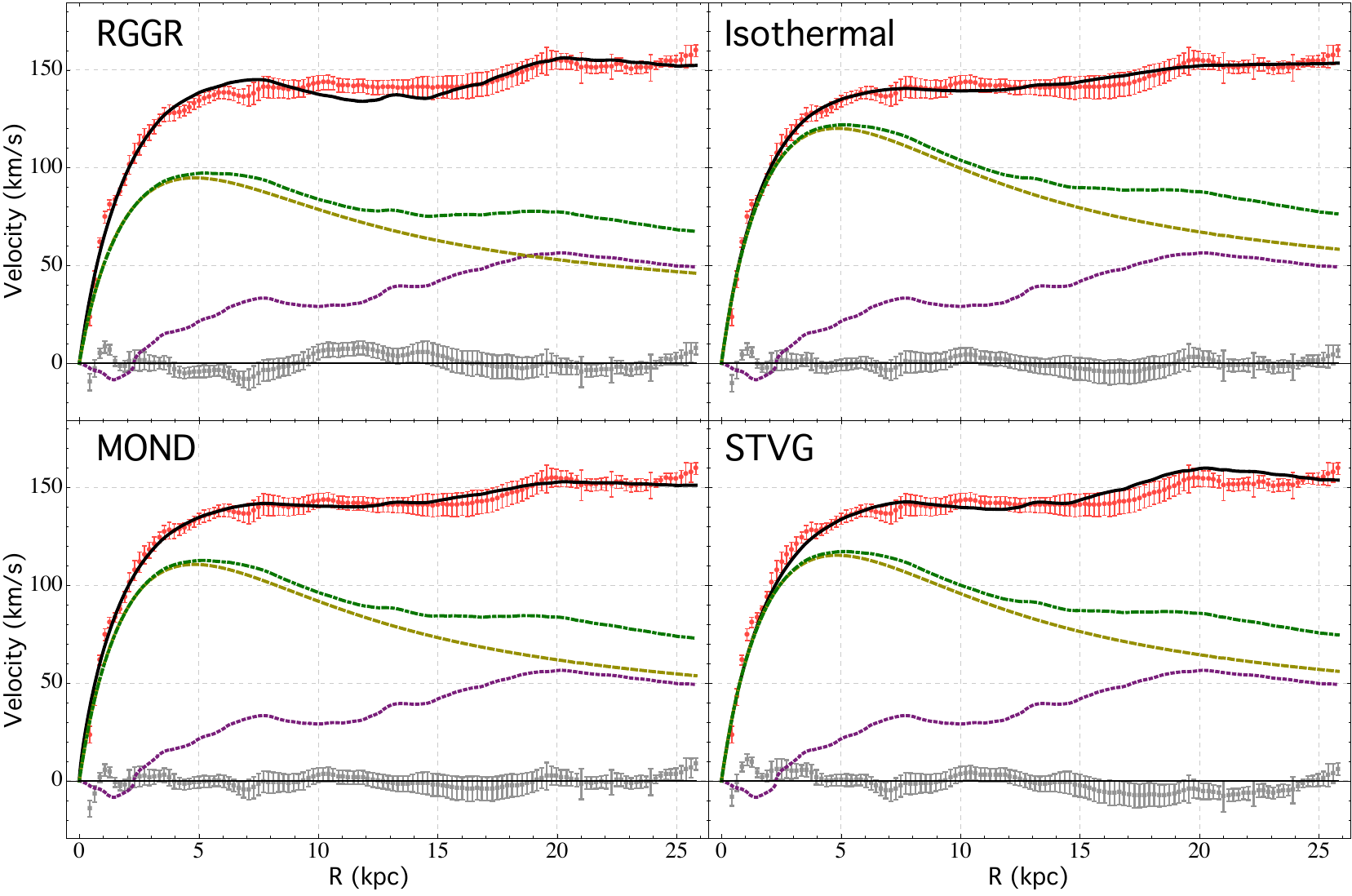} \caption{\label{ngc3621plots} NGC 3621 rotation curve fits. See Fig.(\ref{ddo154plots}) for details.}}

\FIGURE[l]{\includegraphics[width=15cm]{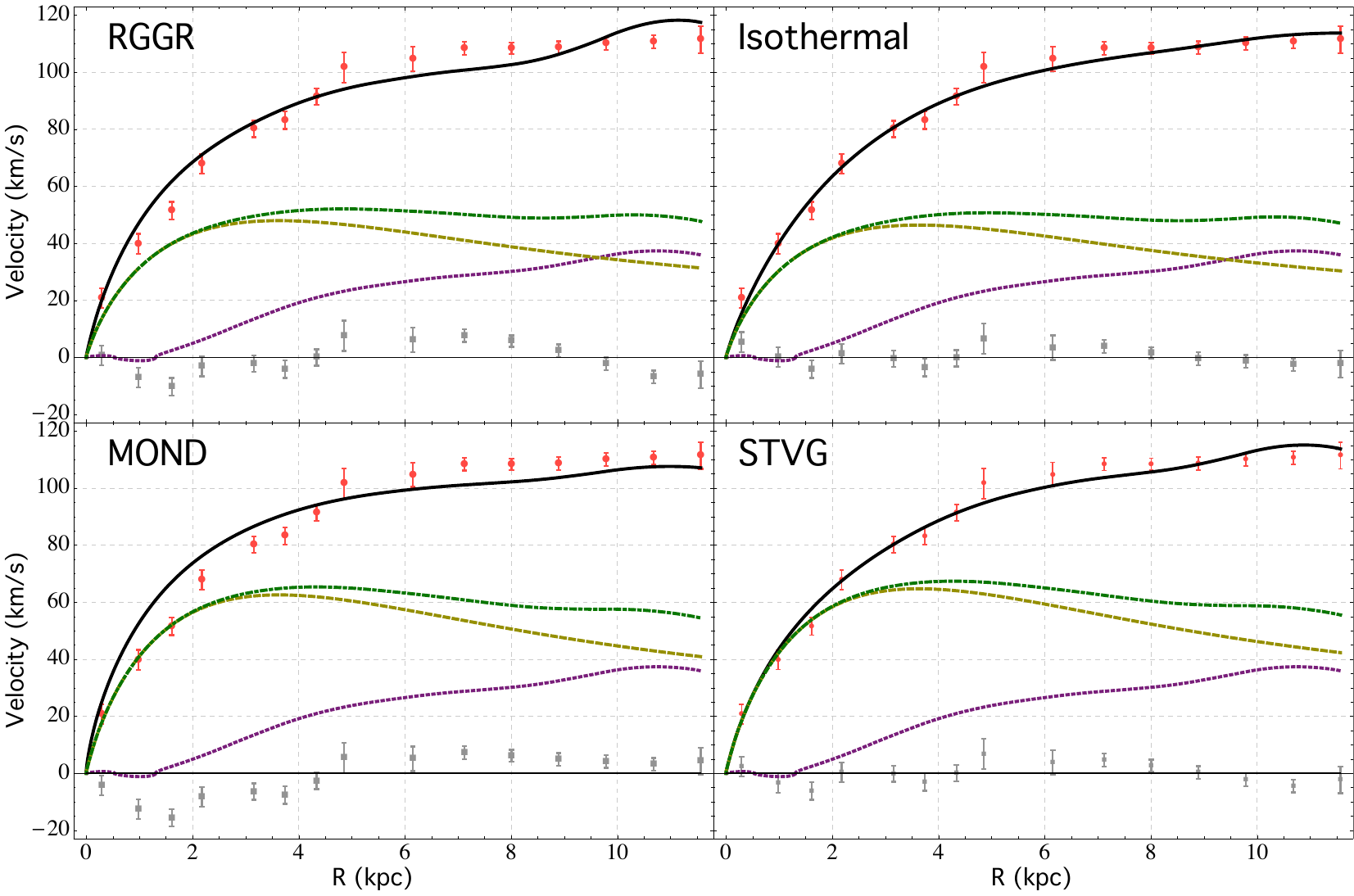} \caption{\label{eso116g12plots} ESO 116-G12 rotation curve fits. See Fig.(\ref{ddo154plots}) for details.}}

\FIGURE[l]{\includegraphics[width=15cm]{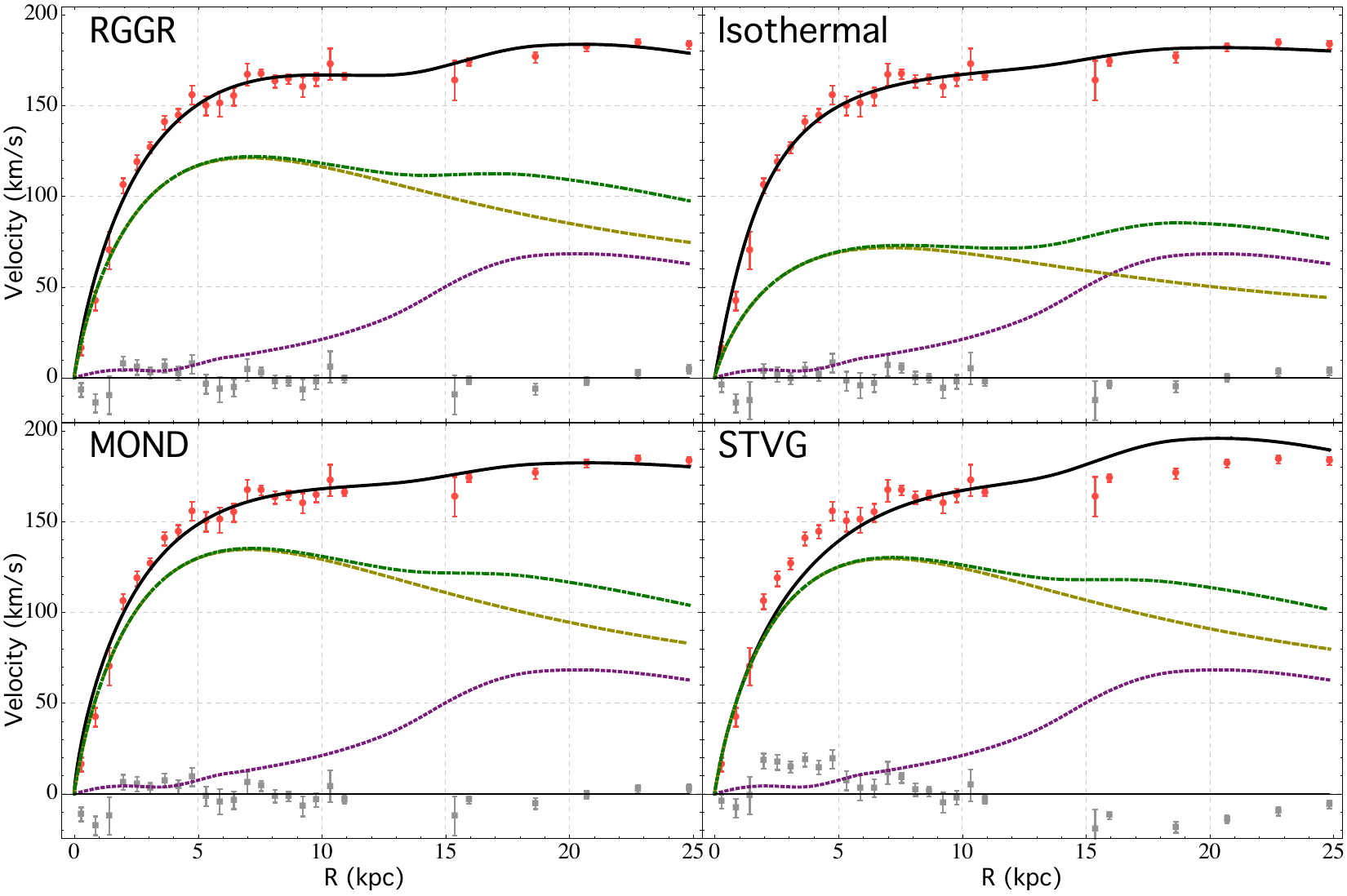} \caption{\label{eso287g13plots} ESO 287-G13 rotation curve fits. See Fig.(\ref{ddo154plots}) for details.}}

\FIGURE[l]{\includegraphics[width=15cm]{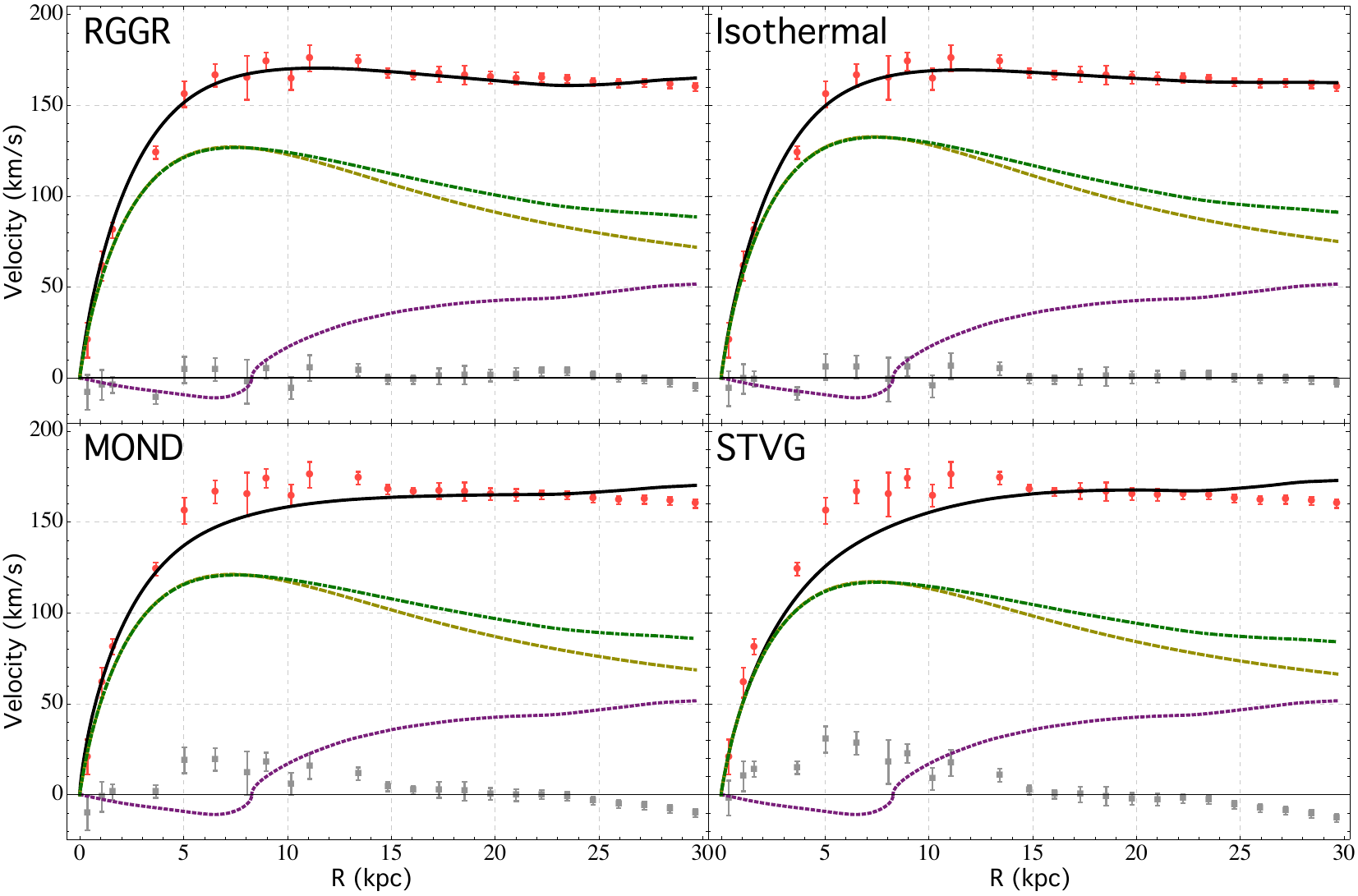} \caption{\label{ngc1090plots} NGC 1090 rotation curve fits. See Fig.(\ref{ddo154plots}) for details.}}

\FIGURE[l]{\includegraphics[width=15cm]{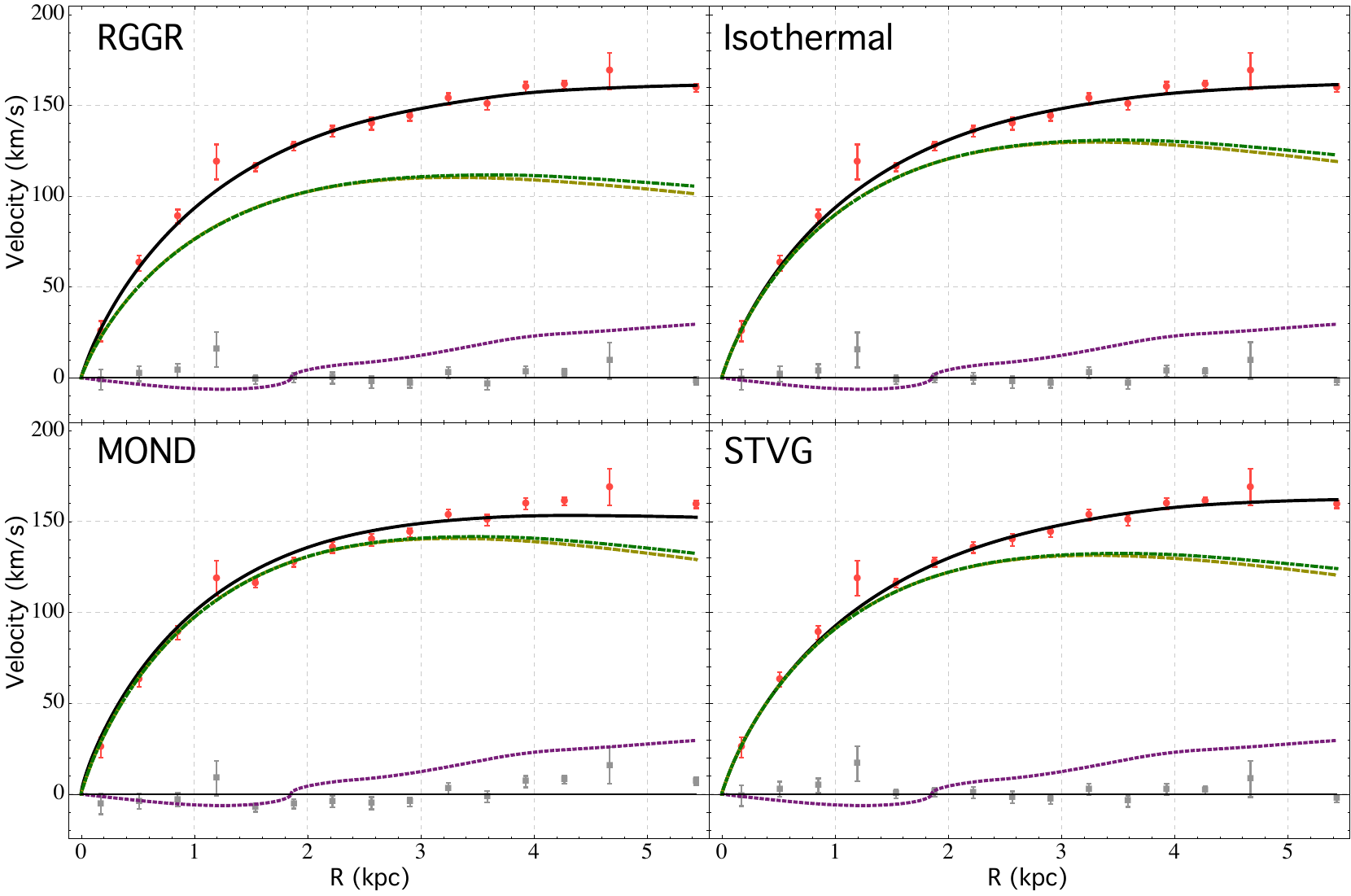} \caption{\label{ngc7339plots} NGC 7339 rotation curve fits. See Fig.(\ref{ddo154plots}) for details.}}

\FIGURE[l]{\includegraphics[width=15cm]{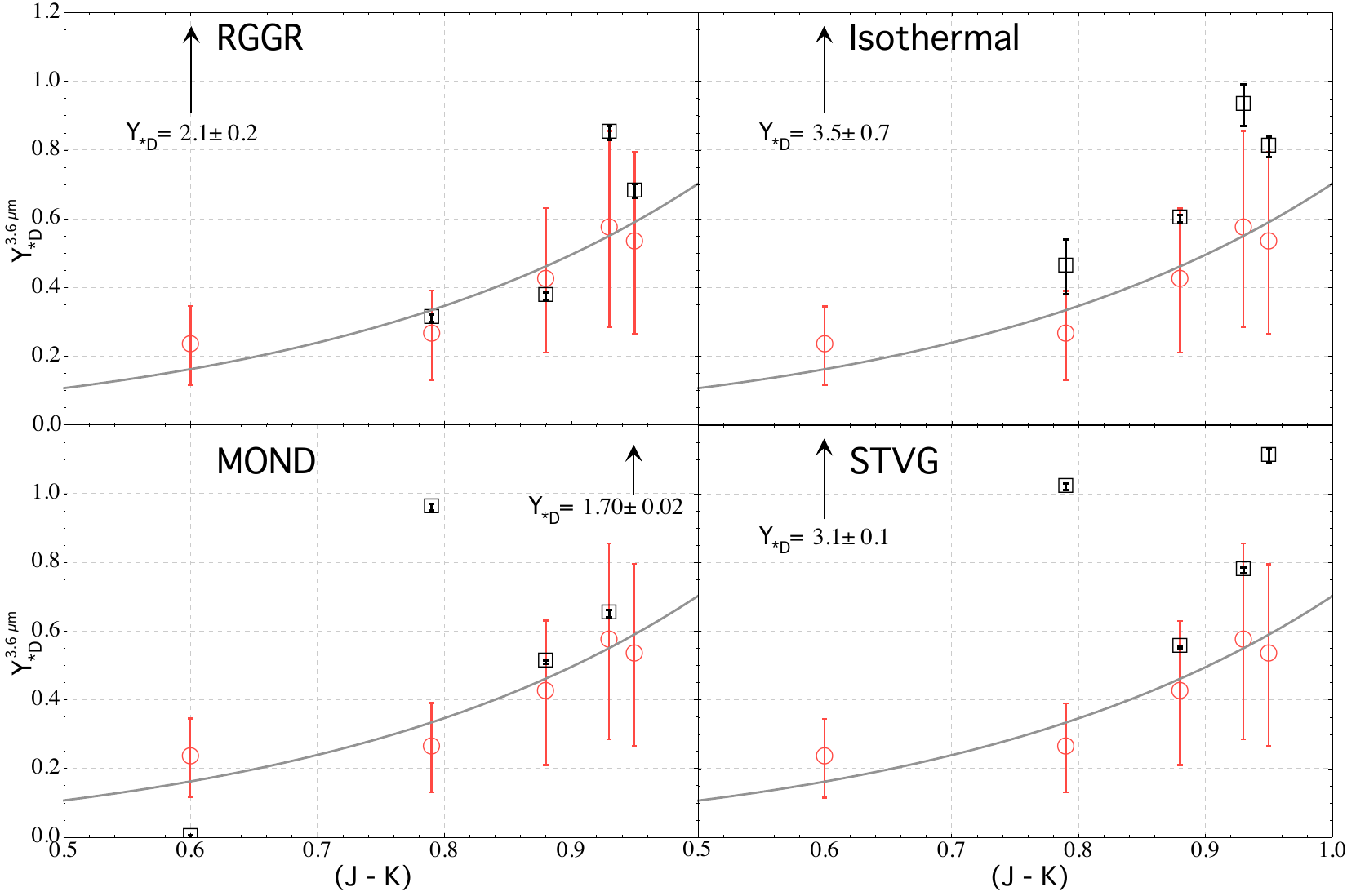} \caption{\label{ColorVsY_b}  { Stellar disk mass-to-light ratio ($Y_{*D}$) in the $3.6  \mu m$ band  as a function of the color $J - K$, for Sample A. The gray curve depicts the $Y_{*D}$ vs. $J-K$ relation according to the Kroupa initial mass function \cite{kroupa, belljong} (see also \cite{ThingsRot}).  The $J-K$ color of each galaxy is determined from the 2MASS Large Galaxy Atlas \cite{2mass}, except for DDO 154 (see notes in \cite{ThingsRot}). Each galactic disk is represented above by an open circle, with a {\it reference} error bar of 50$\%$ of the $Y_{*D}$ value (a bit bigger than the maximum $Y_{*D}$ given by the diet-Salpeter initial mass function \cite{belljong}). The small scatter of the circles around the theoretical line is due to small color gradients present in the stellar disks. The galactic disks depicted above are from, in order of increasing  color: DDO 154, NGC 2403, NGC 3621, NGC 3198, NGC 2841. The black open squares display the $Y_{*D}$ values and their associated errors for each galaxy as inferred from the rotation curve fits for each model.}}}

 
\section{Conclusions and discussions} \label{conclusions}

We  presented a model, motivated by quantum corrections 
to the Einstein-Hilbert action, that introduces small inhomogeneities 
in the gravitational coupling across a galaxy (of about 1 part in 
$10^7$) and can generate galaxy rotation curves in agreement with 
the observational data, without the introduction of dark matter as 
a new kind of matter. Considering the samples of  galaxies evaluated 
in this paper, the quality of the rotation curve fits and the 
associated mass-to-light ratios from our model is a bit lower, or 
about the same, than  the Isothermal profile quality, but with one 
less free parameter. We also compared the results of our model with 
MOND \cite{mond} and STVG\cite{stvg,BrownMoffat}, and at face value 
our model yielded clearly better results.   

Our results can be seen as a next step compared to the 
previous models motivated by renormalization group effects in 
gravity, \cite{Gruni, Reuter:2007de}. Their original analyses 
could only yield a rough estimate on the galaxy rotation curves, 
since they were restricted to modeling a galaxy as a single 
point. Trying to extend this approach to real galaxies, we 
have shown that the phenomenologically optimized proper scale 
for the renormalization group phenomenology is not of a 
geometric type, like the inverse of the distance, but is
proportional to the Newtonian potential with null boundary 
condition at infinity. This setting is shown to be in agreement 
with  both  theoretical expectations and with observations.

The essential feature for the rotation curve fittings is the formula (\ref{vsvg}), which is by itself a remarkably simple formula that provides a very efficient description of galaxy rotation curves. We have shown that  it can be derived from the assumption of a gravitational coupling parameter with a very small departure from the constant $G_0$, such that, in the weak field limit, it depends on the  logarithm of the Newtonian potential, e.g. $G(\Phi_\Newt) = G_0 \[ 1 - \beta \, \ln (\Phi_\Newt / \Phi_0) \]$.  In the last equation, $\beta$ is a positive and very small ($\sim 10^{-7}$) effective parameter that necessarily depends on the distribution of mass of the system. The precise relation between $\beta$ (or $\alpha  \nu$ using the theory of quantum corrections) with the galaxy parameters is yet to be unveiled. To this end, from a phenomenological approach, a larger sample of galaxies would be necessary to find meaningful results. On the other hand, using the Tully-Fisher law, it is not hard to guess that $\beta$, for disk galaxies, should scale with the mass approximately as $\sqrt{M}$, in which $M$ is the baryonic mass of the galaxy. The results for $\alpha$ in tables \ref{resultsA} and \ref{resultsB}, clearly show that it indeed increases with the galactic mass. In order to avoid premature statements, we think that the nature of $\beta$ (or $\alpha$) should be explored in more detail in the forthcoming papers.

In the present paper we have addressed the issue of generating 
galaxies rotation curves by developing and extending the proposal 
of \cite{Gruni}. It turns out that, at least for the sample galaxies we 
dealt with here, one can explain these curves without invoking 
the dark matter concept. Does it mean that we can really be free of
dark matter in constructing a realistic cosmological scenario? 
The answer to this question is probably negative. It is well known 
that there many other issues associated with dark matter, or the 
lack of it, that are necessary to take into account. In particular, 
 the results on the density perturbations and related 
issues such as cosmic microwave background radiation (CMBR) and 
the large scale structure (LSS) data, baryon acoustic oscillations,
big bang nucleosynthesis, gravitational lensing and others (see, 
e.g. \cite{Ross} for a recent review). In all these observational and
experimental issues the standard way of explaining the date is to 
assume the existence of  dark matter. At the same time, one can 
not underestimate the fact that the rotation curves may be explained
without dark matter effect. One can imagine, for instance, the 
scenario with essentially smaller amount of dark matter which 
has slightly different set of properties compared to the usual 
CDM model. In this case the rotation curves will be explained 
by summing up the effect of quantum corrections and the one of 
the dark matter content. The preliminary analysis shows this 
possibility can not be ruled out \cite{AToJFa}. Finally, there is a 
chance to address all mentioned issues trading a large amount 
of cold dark matter content by another one, which can have 
qualitatively other origin, and possibly invoking the quantum 
corrections to the cosmological perturbations spectrum (see, e.g., 
\cite{den, GCC}).

\vspace{.2in}
{\it Note added.}
After the first version of this work was finished, we became aware, through \cite{brownstein}, and contrary to the expectations in \cite{stvg}, that there are small and perhaps sensible differences between the fits of the rotation curves using either the STVG or the MSTG formulations. If precision on the nomenclature and results is at stake, our results on the model developed by Moffat and collaborators apply to the older formulation, namely MSTG.

\vspace{.2in}

\acknowledgments  

We thank  Prof. Erwin  de Blok and Dr. Gianfranco Gentile for providing data on their samples of galaxies.  DCR and PSL thank Dr. Christiane F. Martins for useful comments on galaxy rotation curves.  PSL  thanks CNPq and FAPESP for partial financial support. DCR thanks Mr.  Ronaldo Vieira for useful discussions, the Federal University of Juiz de Fora for hospitality, and FAPESP for financial support. The work of I.Sh. was partially supported by CNPq, FAPEMIG, FAPES and ICTP.


\end{document}